\newif\ifmyrem
\myremtrue

\documentclass[11pt]{article}    
\usepackage{mathtools, amsmath, amssymb, graphics, graphicx, cite, color, float} 
\usepackage{natbib}
\usepackage[labelfont=bf,labelsep=period,justification=raggedright]{caption}

\makeatletter
\renewcommand{\@biblabel}[1]{\quad#1.}
\makeatother

\date{July 15, 2012}

\newenvironment{myitemize}{%
  \begin{itemize}
  \setlength{\parindent}{0pt}
  \setlength{\itemsep}{0pt}
  \setlength{\parskip}{0pt}}{\end{itemize}}

\makeatletter
\def\mylabel#1#2{\label{#1:#2}}%
\def\myref#1#2{%
  \bgroup
  \def\@eq{eq}%
  \def\@fig{fig}%
  \def\@sec{sec}%
  \def\@app{app}%
  \def\@chap{chap}%
  \def\@par{par}%
  \def\@tab{tab}%
  \def\@la{la}%
  \def\@th{th}%
  \def\@def{def}%
  \def\@cor{cor}%
  \def\@alg{alg}%
  \def\@rp{rp}%
  \def\@com{com}%
  \def\@sam{sam}%
  \def\@pro{pro}%
  \def\@dummy{#1}%
  \ifx\@dummy\@eq
    equation~(\ref{#1:#2})%
  \else\ifx\@dummy\@fig
    figure~\ref{#1:#2}%
  \else\ifx\@dummy\@sec
    section~\ref{#1:#2}%
  \else\ifx\@dummy\@app
    appendix~\ref{#1:#2}%
  \else\ifx\@dummy\@chap
    chapter~\ref{#1:#2}%
  \else\ifx\@dummy\@par
    paragraph~\ref{#1:#2}%
    section~\ref{#1:#2}%
  \else\ifx\@dummy\@tab
    table~\ref{#1:#2}%
  \else\ifx\@dummy\@def
    definition~(\ref{#1:#2})%
  \else\ifx\@dummy\@la
    lemma~(\ref{#1:#2})%
  \else\ifx\@dummy\@th
    thesis~(\ref{#1:#2})%
  \else\ifx\@dummy\@cor
    corrolary~(\ref{#1:#2})%
  \else\ifx\@dummy\@alg
    algorithms~(\ref{#1:#2})%
  \else\ifx\@dummy\@rp
    rule~(\ref{#1:#2})%
  \else\ifx\@dummy\@com
    comment~(\ref{#1:#2})%
  \else\ifx\@dummy\@sam
    example~(\ref{#1:#2})%
  \else\ifx\@dummy\@pro
    proposition~(\ref{#1:#2}%
  \else
    (\ref{#1:#2})%
  \fi\fi\fi\fi\fi\fi\fi\fi\fi\fi\fi\fi\fi\fi\fi\fi\egroup}%

\def\Myref#1#2{%
  \bgroup
  \def\@eq{eq}%
  \def\@fig{fig}%
  \def\@sec{sec}%
  \def\@app{app}%
  \def\@chap{chap}%
  \def\@par{par}%
  \def\@tab{tab}%
  \def\@la{la}%
  \def\@th{th}%
  \def\@def{def}%
  \def\@cor{cor}%
  \def\@alg{alg}%
  \def\@rp{rp}%
  \def\@com{com}%
  \def\@sam{sam}%
  \def\@pro{pro}%
  \def\@dummy{#1}%
  \ifx\@dummy\@eq
    Equation~(\ref{#1:#2})%
  \else\ifx\@dummy\@fig
    Figure~\ref{#1:#2}%
  \else\ifx\@dummy\@sec
    Section~\ref{#1:#2}%
  \else\ifx\@dummy\@app
    Appendix~\ref{#1:#2}%
  \else\ifx\@dummy\@chap
    Chapter~\ref{#1:#2}%
  \else\ifx\@dummy\@par
    Paragraph~\ref{#1:#2}%
    Section~\ref{#1:#2}%
  \else\ifx\@dummy\@tab
    Table~\ref{#1:#2}%
  \else\ifx\@dummy\@def
    Definition~(\ref{#1:#2})%
  \else\ifx\@dummy\@la
    Lemma~(\ref{#1:#2})%
  \else\ifx\@dummy\@th
    Thesis~(\ref{#1:#2})%
  \else\ifx\@dummy\@cor
    Corrolary~(\ref{#1:#2})%
  \else\ifx\@dummy\@alg
    Algorithms~(\ref{#1:#2})%
  \else\ifx\@dummy\@rp
    Rule~(\ref{#1:#2})%
  \else\ifx\@dummy\@com
    Comment~(\ref{#1:#2})%
  \else\ifx\@dummy\@sam
    Example~(\ref{#1:#2})%
  \else\ifx\@dummy\@pro
   Proposition~(\ref{#1:#2}%
  \else
    (\ref{#1:#2})%
  \fi\fi\fi\fi\fi\fi\fi\fi\fi\fi\fi\fi\fi\fi\fi\fi\egroup}%
\makeatother

\def \cof{{\sc CoFold}}
\def \cofA{{\sc CoFold-A}}
\def \RNAf{{\sc RNA\-fold}}
\def \RNAfA{{\sc RNA\-fold-A}}
\def \Mfold{{\sc Mfold}}
\def \And{Andronescu}

\def \ss{secondary structure}

\def \eg{\emph{e.g.}}
\def \ie{\emph{i.e.}}
\def \etal{\emph{et al.}}

\def \iviv{\emph{in vivo}}
\def \SI{Supplementary Information}
\def \CRW{{\sc CRW}}
\def \RFAM{{\sc Rfam}}

\begin{document}

\title{\cof: thermodynamic RNA~structure prediction with a kinetic twist}

\author{
Jeff R.~Proctor and Irmtraud M.~Meyer\\
Centre for High-Throughput Biology \& Department of Computer Science and\\
Department of Medical Genetics, University of British Columbia,\\
2125 East Mall, Vancouver, BC, \\
Canada V6T 1Z4, irmtraud.meyer@cantab.net}

\date{July 15, 2012}

\maketitle

\paragraph{Running head:} co-transcriptional RNA folding, RNA structure 
prediction, RNA secondary structures

\bigskip

%
%

\paragraph{Summary:} 

Existing state-of-the-art methods that take a single RNA~sequence and predict
the corresponding RNA~\ss\ are thermodynamic methods. These predict the most
stable RNA~structure, but do not consider the process of structure formation.
We have by now ample experimental and theoretical evidence, however, that
sequences \iviv\ fold while being transcribed and that the process of
structure formation matters. We here present a conceptually new method for
predicting RNA~\ss, called \cof, that combines thermodynamic with kinetic
considerations.  Our method significantly improves the state-of-art in terms
of prediction accuracy, especially for long sequences of more than a thousand
nucleotides length such as ribosomal RNAs.

%
%
%
%




\paragraph{Introduction:}

The primary products of almost all genomes are transcripts, \ie\
RNA~sequences. Their expression is often regulated by RNA~structure which
forms when the transcript interacts with itself via hydrogen-bonds between
complementary nucleotides (G-C, A-U, G-U). These structures regulate
translation, transcription, splicing, RNA~editing and transcript
degradation. To assign a potential functional role to a transcript, it often
suffices to know its RNA~\ss, \ie\ the set of base pairs. As entire
transcriptomes are now routinely sequenced, computational methods that predict
RNA~\ss\ for individual input RNA~sequences play a key role in annotating new
transcripts. This is emphasised by the fact that the majority of mammalian
genomes is transcribed into transcripts of unknown
function~\cite{Mattick2006,Carninci2005} and that experimental techniques for
RNA~structure determination such as X-ray crystallography and NMR remain
costly and slow.

More than three decades of research have been invested into devising methods
that take a single RNA~sequence and predict its RNA~\ss. When homologous
sequences from related species are scarce or not available, non-comparative
methods such as \RNAf~\cite{Zuker1981} and \Mfold~\cite{Zuker2003} provide the
state-of-art in terms of prediction accuracy. They employ an optimisation
strategy that searches the space of potential \ss s for the most stable
structure and depend on hundreds of free energy parameters that have been
initially experimentally determined~\cite{Mathews99} and computationally
tweaked~\cite{Andronescu2007}. Recent attempts at replacing these
thermodynamic parameters by probabilistic ones have lead to a similar or
slightly improved prediction accuracy~\cite{Rivas2012}. All non-comparative
thermodynamic methods, however, show a marked drop in performance accuracy for
increased sequence lengths.

Key experiments~\cite{Boyle1980, Kramer1981, Brehm1983} from the early 1980s
show that structure formation happens co-transcriptionally, \ie\ while the RNA
is being transcribed. Many experiments~\cite{Lewicki1993, Chao1995, Pan1999,
  HeilmanMiller2003, HeilmanMiller2003b, Mahen2005, Adilakshmi2009, Mahen2010,
  Woodson2010} have since substantiated this view. In 1996, Morgan and
Higgs~\cite{Morgan1996} studied the discrepancies between the conserved
RNA~\ss s and the corresponding, predicted minimum free energy (MFE)
structures for long RNA~sequences and concluded that these differences
``cannot simply be put down to errors in the free energy parameters used in
the model''. They hypothesised that these differences may be due to effects of
kinetic folding. These results are complemented by statistical evidence that
structured transcripts not only encode information on the functional
RNA~structure, but also on their co-transcriptional folding
pathway~\cite{Meyer_cf2004}. While there is thus ample evidence that the
process of structure formation matters to the formation of the functional
structure \iviv, it is ignored by thermodynamic methods for RNA~\ss\
prediction.

Several sophisticated computational methods have already been devised that
explicitly mimic the co-transcriptional structure formation \iviv
\cite{Mironov1985, Mironov1993, Gultyaev1995, Flamm:2000rc, Isambert:2000ay,
  Xayaphoummine:2003fc, Xayaphoummine:2005jx, Danilova2006,
  Geis:2008hh}. These folding pathway prediction methods make a range of
simplifying assumptions and approximations of the complex \iviv\ environment.
So far, these methods have only been used to study a few select and typically
short ($\ll$ 1000~nt) sequences and an evaluation of their prediction accuracy
is currently missing.

We here propose a conceptually new method, called \cof, that combines the
benefits of a deterministic, thermodynamic method with kinetic considerations
that capture effects of the structure formation process. For this, we build
upon the state-of-the-art method \RNAf~\cite{Zuker1981} by combining its
thermodynamic energy scores with a scaling function. We train the two free
parameters of \cof\ on a large and diverse data set of 248~sequences and
examine the predictive power of \cof\ on a non-redundant data set of 61~long
sequences ($>$ 1000~nt). \cof\ shows a significant improvement in prediction
accuracy, in particular for long RNA~sequences such as ribosomal RNAs.

\paragraph{Folding long RNA~sequences}

\begin{table}[H]
\centering
\begin{tabular}{l|rrrr}\hline
          & TPR (\%) & FPR (\%) &  PPV (\%) & MCC (\%) \\ \hline
\RNAf\    & 46.30    & 0.0176   &  39.74    & 42.81    \\  
\RNAfA\   & 52.02    & 0.0160   &  44.76    & 48.17    \\
\cof\     & 52.83    & 0.0159   &  45.79    & 49.10    \\
\cofA\    & 57.80    & 0.0145   &  50.06    & 53.70    \\
\end{tabular}
\caption{{\bf Prediction accuracy of \cof\ for base pairs.} 
  The performance accuracy of \cof, \cofA, \RNAf\ and \RNAfA\ 
  for the long data set in terms of true positive rate ($TPR = 100 \cdot TP/(TP + FN)$), 
  false positive rate ($ FPR = 100 \cdot FP/(FP + TN) $),
  positive predictive value ($ PPV = 100 \cdot TP/(TP + FP)$) and Matthew's correlation coefficient 
  ($ MCC = 100 \cdot (TP \cdot TN - FP \cdot FN)/\sqrt{(TP + FP) \cdot (TP + FN) \cdot (TN + FP) \cdot (TN + FN)}$),
  where TP denotes the numbers of true positives, TN the true negatives, 
  FP the false positives and FN the false negatives.
}
\mylabel{tab}{Performance}
\end{table}

We evaluate the prediction accuracy of \cof\ by comparing the \ss\ predicted
by \cof\ to the known reference \ss s for a test set of 61~long sequences
(long data set). We compile this data set by identifying sequences that are
long ($>$ 1000~nt), correspond to biological sequences, have reference
structures that are supported by phylogenetic evidence and are non-redundant
in terms of pairwise percent sequence identify (max 85\%) and evolutionary
distance (\SI, Section~1, \Myref{tab}{Dataset}, \Myref{tab}{Dataset2} and
\Myref{tab}{Dataset3}).  These selection criteria yield a data set of 16S
ribosomal~RNA (rRNA) and 23S~rRNA sequences from archaea, bacteria, eukaryotes
and chloroplasts with an average length of 2397~nt (min 1245~nt, max 3578~nt).

Compared to \RNAf\ which is the state-of-the-art thermodynamic RNA~structure
prediction method, \cof\ predicts 7\% more known base pairs at 6\% higher
specificity than \RNAf\ thereby increasing the Matthew's correlation
coefficient (MCC) by 6\% (MCC (\RNAf) = 42.81\%, MCC (\cof) = 49.10\%)
(\Myref{tab}{Performance}, \SI, Section~2). This improvement in overall
performance accuracy can be attributed to a simultaneous increase of the
positive predictive value (PPV) and the true positive rate (TPR) for almost
all individual sequences (\Myref{fig}{performanceTPRversusPPV}) and a
simultaneous slight decrease of the false positive rate (FPR) (\SI, Section~2
and \Myref{fig}{performanceTPRversusFPR}). Both \RNAf\ and \cof\ employ the
default Turner~1999 free energy parameters~\cite{Mathews99}. Combining \cof\
with the \And~2007 free energy parameters~\cite{Andronescu2007} (\cofA)
increases the sensitivity and specificity by a further 4\% (MCC (\cofA) =
53.70\%). Doing the same with \RNAf\ (\RNAfA) also increases the sensitivity
and specificity with respect to \RNAf, but results in a smaller performance
increase than for \cof\ (MCC (\RNAfA) = 48.17\%, MCC (\cof) =
49.10\%). Whereas \cof\ only depends on two free parameters, the \And~2007
free energy model~\cite{Andronescu2007} comprises 363~free parameters that
were trained using machine learning techniques (\SI, Section~3).

\paragraph{Explicitly capturing the structure formation process}

In order to capture effects of co-transcrip\-tional folding in \cof, we
introduce a scaling function $\gamma(d)$.  This function scales the nominal
energy contribution of any base-pair-like interaction depending on the
distance $d$ of the interaction partners along the sequence (\SI, Section~3
and \Myref{fig}{scalingfunction}). It thereby captures that during the
structure formation process, potential pairing partners in close proximity are
easier to identify than more distant ones. This scaling amounts to a
re-weighing of the structure search space that the structure prediction
algorithm explores. Rather than guiding the structure prediction solely based
on thermodynamic considerations as the state-of-the-art methods \RNAf\ and
\Mfold~\cite{Zuker2003} do, \cof\ thus combines kinetic and thermodynamic
considerations.

The scaling function of \cof\ depends on two free parameters, $\alpha$ and
$\tau$ which have a straightforward interpretation (\SI, Section~3 and
\Myref{fig}{scalingfunction}). Our goal in training the two parameters was to
confirm that parameter training is robust and to ensure that \cof\ can be
applied across a wide range of sequence lengths.

To this end, we compiled an extended data set of 248~sequences (combined data
set) which comprises the 61~long sequences of the long data set and, in
addition, 187~short sequences ($\le 1000$~nt length) that also correspond to
biological sequences whose reference structures are supported by phylogenetic
evidence (\SI, Section~1, \Myref{tab}{Dataset}, \Myref{tab}{Dataset2} and
\Myref{tab}{Dataset3}). The sequences in this combined data set have an
average length of 776~nt (min 110~nt, max 3578~nt). Using twenty trials of
five-fold cross-validation for parameter training, we find that the optimal
prediction accuracy in terms of average MCC is obtained by a combination of
$\alpha$ and $\tau$ values whose strong correlation can be described by a
linear function $\alpha = a \cdot \tau + b$, where $a = 6.1 \cdot 10^{-4} \pm
2 \cdot 10^{-5}$ is the slope and $b = 0.105 \pm 0.016$ the intercept ($R^2 =
98.4$\%) (\SI, Section~4 and \Myref{fig}{linearfit}~(left)). Our
cross-evaluation experiments yield optimal parameter combinations that fall
within or near the 95\% confidence interval around the linear fit, thus
confirming the robustness of parameter training (\SI,
\Myref{fig}{linearfit}~(right)).  We use $\alpha = 0.50$ and $\tau = 640$ in
\cof\ and \cofA\ (\SI, \Myref{fig}{scalingfunction}).

\cof\ and \cofA\ outperform \RNAf\ and \RNAfA\ also for short sequences ($\le\
1000$~nt), although the improvement in terms of MCC is less pronounced than
for long sequences (\SI, \Myref{tab}{Performanceall}). \RNAf\ shows a slight
decrease in prediction accuracy when used with the \And~2007 parameters. The
behaviour of \cof\ is in line with our expectation that the beneficial impact
of modelling co-transcriptional folding decreases for short sequences.

We conclude that \cof\ effectively depends only on one free parameter and that
\cof\ and \cofA\ increase the prediction accuracy for all sequence lengths, in
particular for long sequences ($> 1000$~nt).

\paragraph{Capturing structure formation yields improved structures of similar free energies}


In order to examine if capturing the effects of co-transcriptional folding
significantly changes the free energies of the predicted structures, we
calculated the free energies of the structures predicted by \cof, \cofA\ and
\RNAfA\ and compared them to the free energies of the corresponding structures
predicted by \RNAf. To ensure consistency, we used the Turner~1999 energy
parameters to calculate the energies of all predicted RNA~structures.

The structures predicted by \cof\ for the long data set differ on average by
2\% from the respective free energies of the corresponding structures
predicted by \RNAf\ and the distribution of relative energy differences is
comparatively tight (stdev = 1.0\%, min = 0.2\%, max = 4.4\%)
(\Myref{fig}{deltaGdistributions_long} and \SI,
\Myref{tab}{deltaGdistributions}). Combining \cof\ and \RNAf\ with the
\And~2007 energy parameters significantly increases the average free energy
difference (5\% (\RNAfA), 7\% (\cofA)), broadens the distributions
(stdev(\RNAfA) = 1.9\%, stdev(\cofA) = 2.4\%) and leads to higher maximum
energy differences (max(\RNAfA) = 11.1\%, max(\cofA) = 13.1\%).  For short
sequences, these differences are even more pronounced (\SI,
\Myref{tab}{deltaGdistributions}).

Most importantly, a large energy difference with respect to the free energy of
the structure predicted by \RNAf\ does not imply an increased prediction
accuracy, neither for short nor long sequences and for none of the prediction
programs (\SI, \Myref{fig}{deltadeltaGversusdeltaMCCfits} and
\Myref{tab}{deltadeltaGversusdeltaMCCfits}).

To summarise, \cof\ significantly increases the prediction accuracy without
significantly altering the free energies of the structures that \RNAf\ would
predict for the same input sequences.

\paragraph{Folding ribosomal RNAs}




23S~ribosomal RNAs are the longest sequences of the long data set with an
average length of 3069~nt (min 2882~nt, max 3578~nt) and are thus some of the
most challenging RNA~structures to predict. Using \cof\ and \cofA, we increase
their prediction accuracy in terms of MCC w.r.t.\ \RNAf\ on average by 8\% and
12\%, respectively. \Myref{fig}{23S} shows, for the 23S~rRNA of the
gamma-proteobacteria \emph{Pseudomonas aeruginosa}, how the RNA~structure
predicted by \cofA\ compares to that predicted by \RNAf.  The most apparent
differences are that \RNAf\ predicts many incorrect mid- and long-range base
pairs (red arcs spanning more than 100~nt) and that almost all of these
disappear with \cofA. In addition, \cofA\ adds many correct mid- and
long-range base pairs (blue arcs), see in particular those spanning almost the
entire sequence.  Overall, \cofA\ increases the MCC of \RNAf\ from 43\% to
58\%. This 15\% rise in performance accuracy is due to a significant increase
of the true positive rate (45\% $\rightarrow$~61\%) and an equally significant
increase of the positive predictive value (41\% $\rightarrow$~56\%). This is
in-line with is the typical behaviour seen for \cof\
(\Myref{fig}{performanceTPRversusPPV} and \SI,
\Myref{fig}{performanceTPRversusFPR}). The false positive rate for both
prediction methods remains low at 0.01\%.



We also investigated the performance for the 16S~ribosomal RNAs in greater
detail.  With an average length of 1550~nt (min 1245~nt, max 1799~nt), these
are significantly shorter than the 23S~rRNAs in our long data set, but still
considerably longer than the average test sequence on which thermodynamic
prediction methods are typically benchmarked. \Myref{fig}{16S} shows the
improvements in prediction accuracy for the 16S~rRNA of the freshwater algae
\emph{Cryptomonas sp.}  (species unknown).  This ribosomal sequence is 1493~nt
long. \cofA\ improves the prediction accuracy of \RNAf\ from an MCC of 32\% to
73\%. This 41\% improvement in performance accuracy is achieved by
significantly reducing the number of erroneously predicted mid- to long-range
base pairs (red arcs spanning more than 100~nt) while simultaneously
increasing the number of correctly predicted base pairs in wide distance range
(blue arcs). This is reflected by the simultaneous increase of the true
positive rate (33\% $\rightarrow$~77\%) and the positive predictive value
(30\% $\rightarrow$~69\%) which, in this example, is also accompanied by a
slight reduction of the false positive rate (0.03\% $\rightarrow$~0.01\%).

As neither \cof\ nor \RNAf\ are technically capable of predicting
pseudo-knotted features, the pseudo-knotted reference structures of the
23S~rRNA and the 16S~rRNA cannot be predicted with perfect accuracy (see
orange arcs in \Myref{fig}{23S} and \Myref{fig}{16S}).

\paragraph{Discussion}

Our results show that the state-of-the-art in non-comparative RNA~\ss\
prediction can be significantly improved by capturing effects of the structure
formation process. To this end, we introduce a conceptually new RNA~\ss\
prediction method called \cof\ which judges the reachability of potential
pairing partners during co-transcriptional structure formation via a scaling
function. We reliably train the two free parameters of \cof\ using a large and
diverse data set of 248~sequences. This scaling function effectively depends
on only one free parameter which has a straightforward interpretation as it
determines how the reachability declines as function of the nucleotide
distance. Without altering the free energy parameters of the underlying
thermodynamic model, \cof\ thereby guides the structure prediction process by
a combination of thermodynamic and kinetic considerations. It thereby arrives
at significantly more accurate structure predictions, in particular for long
sequences ($>$ 1000~nt) such as ribosomal RNAs.  The significance of the
improvement in prediction accuracy is underlined by the improvement in
sensitivity and specificity for the individual sequences.  Most importantly,
this improvement is gained without significantly shifting the free energies of
the predicted RNA~structures.  We thereby confirm Morgan and
Higgs~\cite{Morgan1996} who hypothesised that discrepancies between the
functional RNA~\ss\ and the corresponding minimum free energy structures
predicted by thermodynamic methods such as \RNAf\ are not due to errors of the
underlying free energy parameters, but due to a lack of modelling the effects
of kinetic structure formation.

Many sophisticated experiments~\cite{Boyle1980, Kramer1981, Brehm1983,
  Lewicki1993, Chao1995, Pan1999, HeilmanMiller2003, HeilmanMiller2003b,
  Mahen2005, Adilakshmi2009, Mahen2010, Woodson2010} paint a dauntingly
complex picture of co-transcriptional structure formation \iviv\ which can
depend on a multitude of factors ranging from the speed of transcription and
the variation thereof, to a range of carefully orchestrated \emph{cis} and
\emph{trans} interactions with a variety of other molecules.

Several sophisticated folding pathway prediction methods have already been
devised that explicitly mimic the co-transcriptional structure formation \iviv
\cite{Mironov1985, Mironov1993, Gultyaev1995, Flamm:2000rc, Isambert:2000ay,
  Xayaphoummine:2003fc, Xayaphoummine:2005jx, Danilova2006, Geis:2008hh}.
Even though these methods need to make a range of simplifying assumptions to
approximate the complex \iviv\ environment and have so far been been only used
to investigate a few select and typically short ($\ll$ 1000~nt) sequences,
these methods have already allowed us to gain valuable and detailed insight
into co-transcriptional folding pathways~\cite{Isambert:2000ay,
  Schoemaker2006}.

By proposing a conceptually new approach to RNA~\ss\ prediction, \cof, we show
that it is possible to capture effects of the structure formation process in
deterministic thermodynamic methods and that the benefits of doing so are
significant, both in terms of prediction accuracy and insight gained.  This
finding is not too surprising given that any co-transcriptionally emerging
RNA~transcript \iviv\ needs to find a way to actually reach the functionally
relevant RNA~structure. Although \cof\ only constitutes the first attempt at
explicitly capturing the effects of co-transcriptional folding in a
thermodynamic RNA~\ss\ prediction program, we hope that our results will
inspire a new generation of these methods that explicitly capture aspects of
the structure formation process \iviv.  Recent developments in experimental
techniques~\cite{Adilakshmi2009, Woodson2010} will no doubt significantly
contribute to our understanding of structure formation \iviv. We are thus
looking forward to joint projects between the experimental and theoretical
RNA~structure community.

\paragraph{Methods summary}

The algorithm of \cof\ and the scoring function are described in detail in the
\SI. \cof\ is publicly available on the Internet via a web-server at
http://www.e-rna.org/cofold.


\bibliographystyle{naturemag}




\paragraph{Acknowledgements} This project was supported by grants to
I.M.M. from the Natural Sciences and Engineering Research Council (NSERC) of
Canada and from the Canada Foundation for Innovation (CFI).  J.R.P. holds an
Alexander Graham Bell Canada Graduate Scholarship from NSERC, with additional
funding from the CIHR/MSFHR Bioinformatics Training Program at the University
of British Columbia. CIHR are the Canadian Institutes of Health Research and
MSFHR is the Michael Smith Foundation for Health Research in Canada.


\paragraph{Author Contributions} Both authors were involved in every aspect of
the research. J.R.P. programmed \cof.


\paragraph{Author Information} 

Correspondence and requests for material should be addressed to I.M.M.\
(irmtraud.meyer@cantab.net).

\newpage

\section*{Figures}

\begin{figure}[H]
  \begin{center}
    \includegraphics[width=8cm]{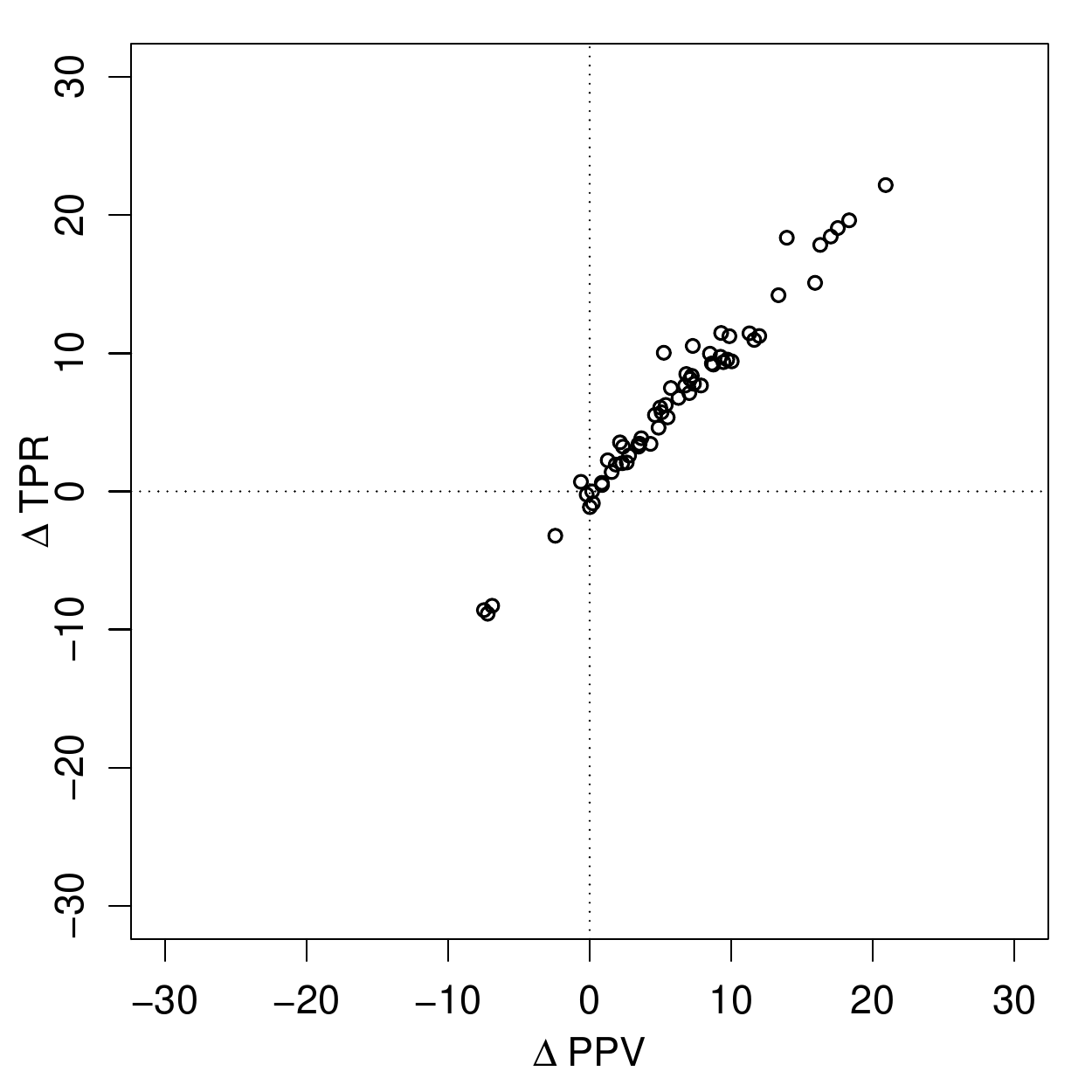}
  \end{center}
  \caption{ {\bf Changes in prediction accuracy for the structures predicted
      by \cof\ for individual sequences.}  We report the prediction accuracy
    for base pairs of the long data set in terms of absolute changes of the
    true positive rate ($TPR = 100 \cdot TP/(TP + FN)$) and the positive
    predictive value ($ PPV = 100 \cdot TP/(TP + FP)$) by comparing the
    prediction accuracy of the structures predicted by \cof\ to those
    predicted by \RNAf. TP denotes the numbers of true positives, TN the true
    negatives, FP the false positives and FN the false negatives, see \SI\ 
    Section~2 for detailed definitions.}
\mylabel{fig}{performanceTPRversusPPV}
\end{figure}

\begin{figure}[H]
   \begin{center}
     \includegraphics[width=8cm]{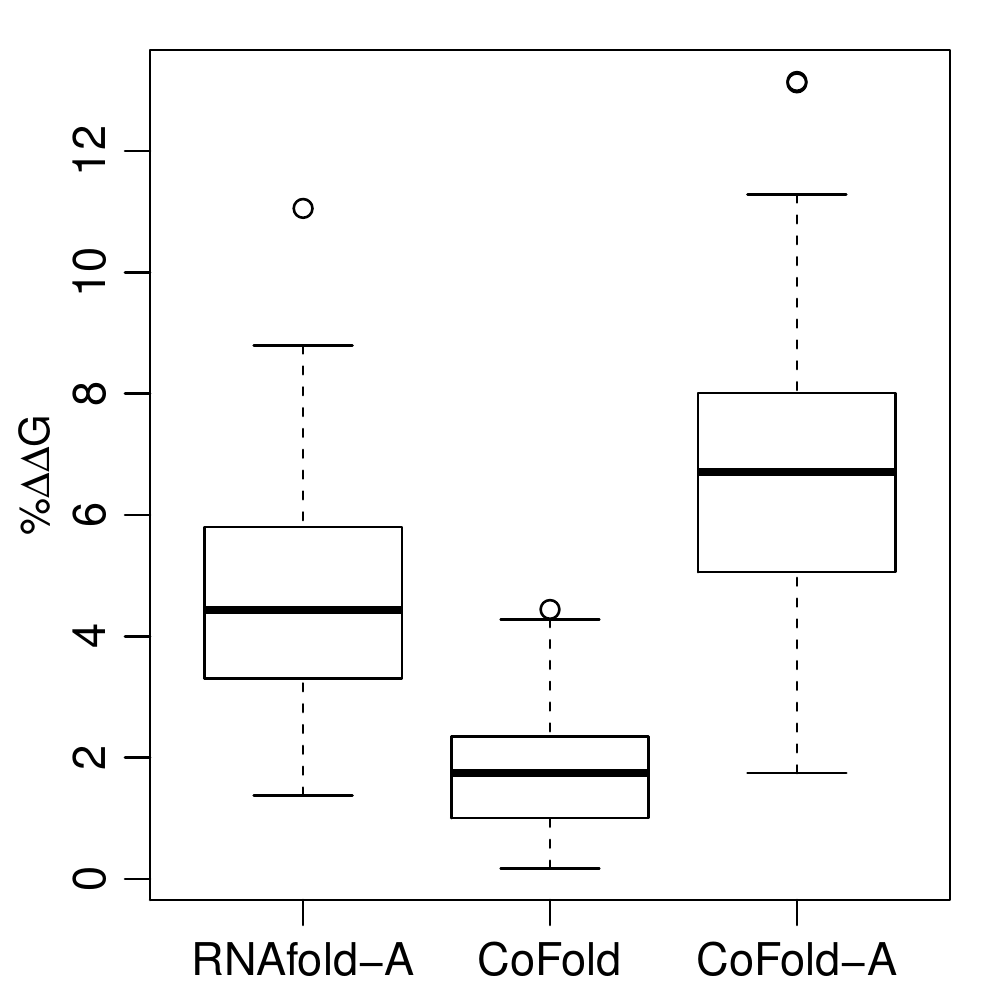}
   \end{center}
  \caption{{\bf Relative free energy differences of the predicted structures
      w.r.t.\ the MFE~structures predicted by \RNAf.}  Summary of three
    distributions for the long data set showing the relative free energy
    differences of the RNA~structures predicted by \RNAfA\ w.r.t. the
    MFE~structures predicted by \RNAf\ for the same sequence (left), of the
    RNA~structures predicted by \cof\ w.r.t.\ the MFE~structures predicted by
    \RNAf\ (middle) and of the RNA~structures predicted by \cofA\ w.r.t\ the
    MFE~structures predicted by \RNAfA\ (right). The free energies of all
    structures are calculated using the Turner~1999 energy parameters. For
    each of the three distributions, the dark horizontal line indicates the
    average, the box indicates the 1st to the 3rd quartile, and the dotted
    lines indicate minimum and maximum values. Circles indicate outliers which
    are not included in the calculation of the average value.}
  \mylabel{fig}{deltaGdistributions_long}
\end{figure}

\begin{figure}[H]
   \begin{center}
     \includegraphics[width=16cm]{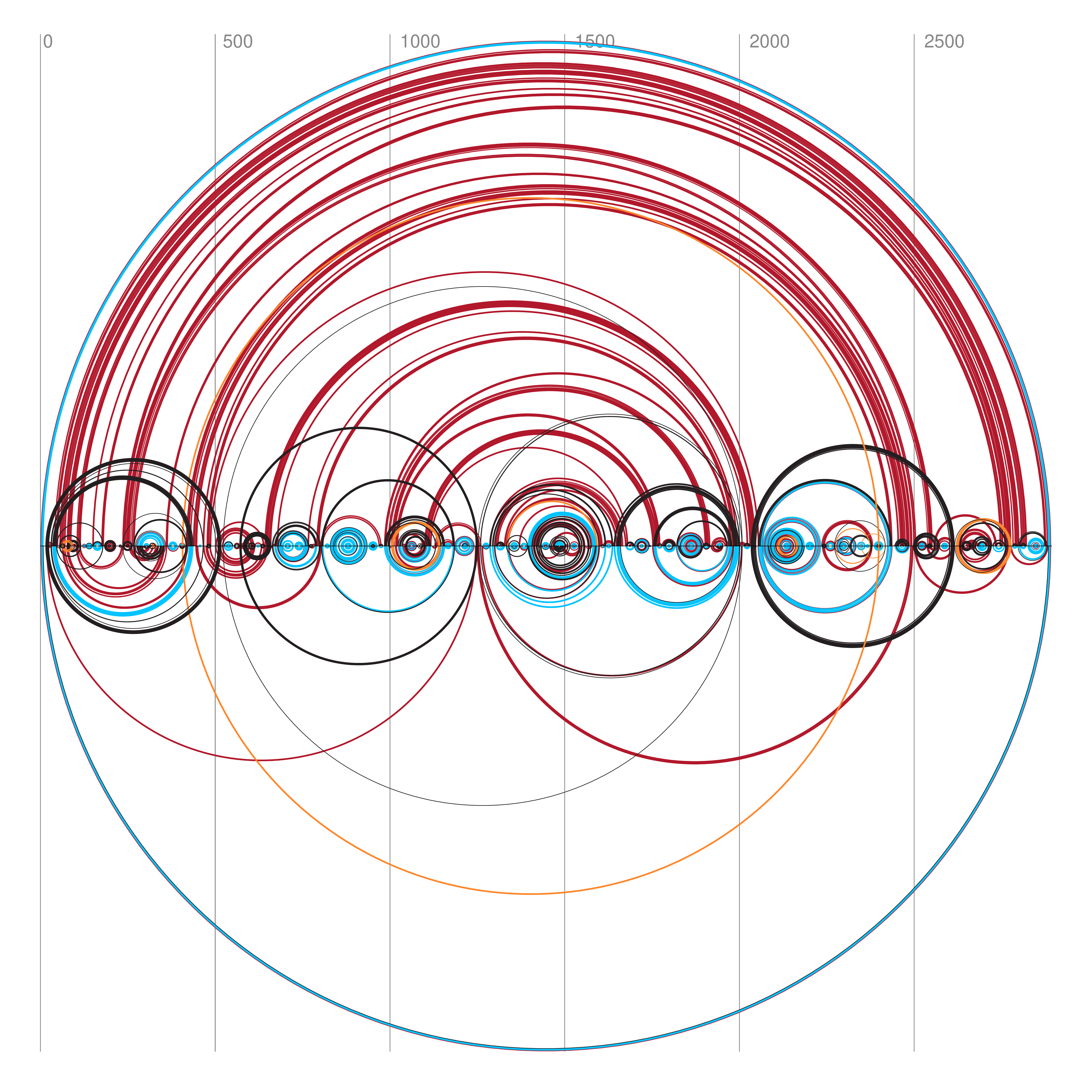}
   \end{center}
  \caption{{\bf RNA~\ss s predicted by \cofA\ and \RNAf\ for the 23S~rRNA of
      the gamma-proteobacteria \emph{Pseudomonas aeruginosa}.}  The horizontal
    line corresponds to the RNA~sequence of 2893~nt length.  The structure
    predicted by \RNAf\ is shown above the horizontal line, the one predicted
    by \cofA\ below. Each arc corresponds to a base-pair between the two
    corresponding positions along the sequence.  Blue arcs correspond to
    correctly predicted base pairs (true positives), red arcs to incorrectly
    predicted base pairs (false positives) and black arcs to base pairs that
    are part of the reference structure, but missing from the prediction
    (false negatives). Orange arcs indicate base pairs of the reference
    structure that render it pseudo-knotted. Figure made with {\sc
     R-chie}~\cite{Lai2012}.}
  \mylabel{fig}{23S}
\end{figure}

\begin{figure}[H]
   \begin{center}
     \includegraphics[width=16cm]{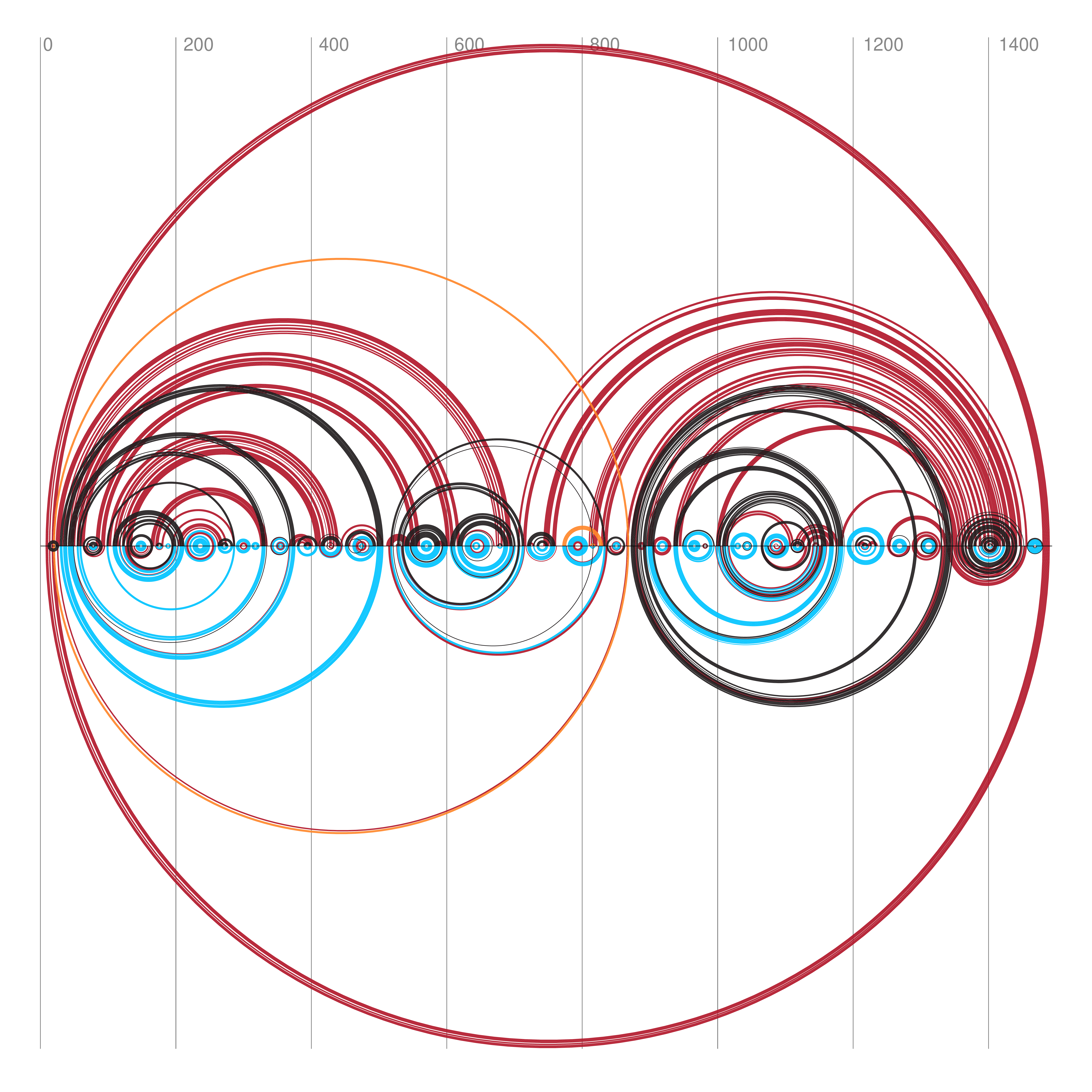}
   \end{center}
  \caption{{\bf RNA~\ss s predicted by \cofA\ and \RNAf\ for the 16S~rRNA of
      the freshwater algae \emph{Cryptomonas sp.}.} The horizontal line
    corresponds to the RNA sequence of 1493~nt length.  The structure
    predicted by \RNAf\ is shown above the horizontal line, the one predicted
    by \cofA\ below. Each arc corresponds to a base-pair between the two
    corresponding positions along the sequence.  Blue arcs correspond to
    correctly predicted base pairs (true positives), red arcs to incorrectly
    predicted base pairs (false positives) and black arcs to base pairs that
    are part of the reference structure, but missing from the prediction
    (false negatives). Orange arcs indicate base pairs of the reference
    structure that render it pseudo-knotted. Figure made with {\sc
      R-chie}~\cite{Lai2012}.}
  \mylabel{fig}{16S}
\end{figure}

\vfill

\pagebreak

\setcounter{page}{1}


\begin{center}

\Large{{\bf \SI}}

\bigskip

\Large{\cof: thermodynamic RNA~structure prediction with a kinetic twist}

\bigskip

\large{
Jeff R.~Proctor and Irmtraud M.~Meyer\\
Centre for High-Throughput Biology \& Department of Computer Science and\\
Department of Medical Genetics, University of British Columbia,\\
2125 East Mall, Vancouver, BC, \\
Canada V6T 1Z4, irmtraud.meyer@cantab.net}

\medskip

\date{July 15, 2012}      
\end{center}

\bigskip

\subsection*{Methods}

\paragraph{(1) Compilation of the long and combined data sets}

The long data set consists of 16S and 23S ribosomal RNAs only.  16S and 23S
multiple-sequence alignments for bacteria, eukaryotes, archaea and
chloroplasts were retrieved from the Comparative RNA Web site
(\CRW)~\cite{Cannone2002}.  Because no consensus RNA~structure is provided for
each alignment, we projected the individual structures onto the alignment. The
structure with the lowest mismatch score (defined in
\Myref{eq}{structmismatch}) was chosen as the consensus RNA~structure for each
alignment.  For the calculation of the mismatch score, base pairs with a gap
in one base position, and a non-gap in the other are considered one-sided
gaps. Base pairs with gaps on both sides are considered two-sided gaps.
Non-canonical pairs are those other than G-C, A-U, G-U. The length of the
alignment $A$ is denoted by $|A|$.

\begin{equation}
    \mylabel{eq}{structmismatch}
    \text{mismatch} := \frac{\sum_{seq \in A} \left(2 \cdot (\text{\# one-sided gaps}) + (\text{\# two-sided gaps}) +
(\text{\# non-canonical pairs})  \right)}{|A|}
\end{equation}

Sequences with large indels, many ambiguous nucleotides, or a poor fit to the
consensus RNA~structure were removed from the alignment. Unpaired regions of
the alignment were realigned using {\sc MUSCLE}~\cite{Edgar2004}.  Individual
sequences were extracted from each resulting alignment such that no pair of
extracted sequences have a pairwise percent sequence identity greater than an
alignment-specific threshold. The exact threshold varies to ensure no
biological class or evolutionary clade is over-represented in the long data
set (max $85$\%, \Myref{tab}{Dataset2}).  For each extracted sequence, the
consensus alignment structure was projected onto the sequence by removing base
pairs at gap positions, and removing any non-canonical base pairs. The
resulting 61~sequences have a mean sequence length of $2397$~nt and constitute
the long data set (\Myref{tab}{Dataset}, \Myref{tab}{Dataset2} and
\Myref{tab}{Dataset3}).

The combined data set was constructed primarily for robustness of parameter
training, and contains \RFAM\ sequences from a wide variety of biological
classes~\cite{Griffiths-Jones2005}. \RFAM\ alignments were chosen according to
the following criteria:

\begin{myitemize}
    \item mean sequence length greater than $115$
    \item covariation (defined in \Myref{eq}{covariationeq}) greater than $0.18$ 
    \item minimum of $5$ sequences
    \item mean percentage of canonical base pairs greater than $80$\%
    \item diverse biological classes and evolutionary clades
\end{myitemize}

Sequences were extracted from the \RFAM\ alignments using the same protocol as
for the \CRW\ alignments described above.  Specifically, no pair of sequences
extracted from the same alignment share a pairwise percent sequence identity
above an alignment-specific threshold (max $85$\%, \Myref{tab}{Dataset2}).
Consensus RNA~structures were projected onto individual sequences by removing
base pairs at gap positions, and by removing any non-canonical base pairs. The
mean sequence length of the resulting $187$ \RFAM\ sequences is $247$~nt, and
the combined dataset has an average sequence length of $778$~nt
(\Myref{tab}{Dataset}), see \Myref{tab}{Dataset2} for a description of
biological class and sequence extraction details, and \Myref{tab}{Dataset3}
for alignment quality metrics.

For a given multiple-sequence alignment, the covariation is defined as:
\begin{equation}
    \mylabel{eq}{covariationeq}
\text{covariation} := \frac{\sum^M_{a=1,b=1,a<b} \left( \sum_{S_{ij}} (\Pi^{ab}_{ij}H(a_i a_j, b_i b_j) - \Omega^{ab}_{ij}H(a_i a_j, b_i b_j)) \right) / (|S_{ij}|) }{\binom{M}{2}}
\end{equation}

where
\begin{myitemize}
\item $S_{ij}$ is the set of base pairs $i$ and $j$ in the consensus secondary structure.
\item $M$ is the number of sequences in the alignment.
\item $H(a_i a_j, b_i b_j)$ is the Hamming distance between the strings $a_i a_j$ and $b_i b_j$.
\item $\Pi^{ab}_{ij}$ is an indicator function such that if $a_i$ and $a_j$ can form a canonical base-pair, and $b_i$ and $b_j$ can also form a canonical base-pair, $\Pi^{ab}_{ij} = 1$ (otherwise $\Pi^{ab}_{ij} = 0$).
\item $\Omega^{ab}_{ij}$ is an indicator function such that if $a_i$ and $a_j$ and/or $b_i$ and $b_j$ cannot for a canonical base-pair, $\Omega^{ab}_{ij} = 1$ (otherwise $\Omega^{ab}_{ij} = 0$).
\end{myitemize}

\paragraph{(2) Definition of performance metrics}

Structure prediction accuracy is measured on a base pair level. True positives
(TP) are correctly predicted base pairs. False positives (FP) are incorrectly
predicted base pairs that are not part of the reference structure. True
negatives (TN) are hypothetical base pairs that are not predicted, nor part of
the reference structure. False negatives (FN) are reference base pairs missed
by the prediction. Performance metrics for true positive rate (TPR), false
positive rate (FPR), positive predictive value (PPV), and Matthew's
correlation coefficient (MCC) are defined as follows:

\begin{equation}
    TPR = 100 \cdot \frac{TP}{TP + FN}
\end{equation}

\begin{equation}
    \Delta TPR = TPR_{\text{CoFold}} - TPR_{\text{RNAfold}}
\end{equation}

\begin{equation}
    FPR = 100 \cdot \frac{FP}{FP + TN}
\end{equation}

\begin{equation}
    \Delta FPR = FPR_{\text{CoFold}} - FPR_{\text{RNAfold}}
\end{equation}

\begin{equation}
    PPV = 100 \cdot \frac{TP}{TP + FP}
\end{equation}

\begin{equation}
    \Delta PPV = PPV_{\text{CoFold}} - PPV_{\text{RNAfold}}
\end{equation}

\begin{equation}
    MCC = 100 \cdot \frac{TP \times TN - FP \times FN}
    {\sqrt{(TP + FP) \times (TP + FN) \times (TN + FP) \times (TN + FN)}}
\end{equation}

\paragraph{(3) Incorporating co-transcriptional folding into the prediction algorithm of \cof}

The Nussinov algorithm \cite{Nussinov1980} was one of the first attempts at
RNA \ss\ prediction. It is a dynamic programming method that efficiently
calculates the \ss\ with the largest number of base pairs in $O(L^{3})$ time,
where $L$ denotes the length of the input sequence.  The algorithm solves the
problem recursively by determining the optimal structure for sub-sequences,
and using these solutions to derive optimal structures for successively larger
subsequences. The output structure is the optimal solution for the full
sequence.  This algorithm, however, has several shortcomings. First, base
pairs vary in stability; for example, G-C pairs are energetically more
favourable than A-U pairs. The Nussinov algorithm weights all pairs equally.
Second, The stability of a base pair depends highly on its neighbouring base
pairs due to so-called stacking interactions between adjacent pairs, and this
contextual effect is ignored by the algorithm.

The Zuker-Stiegler algorithm \cite{Zuker1981} is an advancement of the
Nussinov algorithm.  Rather than predicting the structure with the greatest
number of pairs, the Zuker-Stiegler algorithm predicts the most
thermodynamically favourable (and pseudo-knot free) RNA~structure according to
a set of free energy parameters. This structure is also called the
minimum-free-energy (MFE) structure. The algorithm assigns a sequence-specific
free energy value to various structural building blocks, such as stacking
interactions between pairs of adjacent base pairs, unpaired nucleotides, and
hairpin loops.  The algorithm utilises dynamic programming similarly to the
Nussinov algorithm, but calculates two energy values for all subsequences
$S_{i,j}$ of a given input sequence $S$, where $1 <= i < j <= L$:

\begin{itemize}
    \item $C_{i,j}$: minimum free energy of subsequence $S_{i,j}$ given nucleotides i and j form a base pair
    \item $FML_{i,j}$: minimum free energy of subsequence $S_{i,j}$
\end{itemize}

\begin{equation}
    \mylabel{eq}{cijeq}
        C_{i,j} = min \begin{dcases*}
                hairpin_{i,j} &  open a helix with hairpin loop \\
                min_{i<p<q<j} \{C_{p,q} + Stack_{(i,j),(p,q)}\} & stack, bulge or internal loop \\
                min_{k,l \in {1,2}} \{ FML_{i+k, j-l} + dangle \} & open a helix with nested substructure \\
          \end{dcases*} 
\end{equation}

\begin{equation}
    \mylabel{eq}{fmlijeq}
        {FML}_{i,j} = min \begin{dcases*}
                 min_{i<k<j} \{FML_{i,k} + FML_{k+1,j}\} & branched structures \\
                 min_{k,l \in {0,1}} \{C_{i+k, j-l} + dangle\} & close a helix \\
                 FML_{i+1,j} + E_{unpaired} & unpaired nucleotide \\
                 FML_{i,j-1} + E_{unpaired} & unpaired nucleotide \\
         \end{dcases*}
\end{equation}

$C_{i,j}$ and $FML_{i,j}$ are calculated for each subsequence $S_{i,j}$ as the
minimum of a well-defined set of rules (\Myref{eq}{cijeq},
\Myref{eq}{fmlijeq}).  The minimum free energy can be retrieved from the value
at $FML_{1,L}$, where $L$ denotes the length of the input sequence. The
corresponding MFE~structure is retrieved by backtracking through the $C_{i,j}$
and $FML_{i,j}$ matrices.

The Zuker-Stiegler algorithm requires a large set of thermodynamic parameters.
In 1999, the Turner group published one such model, which included a
combination of experimentally measured energies and estimated values
\cite{Mathews99}. This parameter set (called Turner~1999 parameter set) is
widely used by many state-of-the-art tools, including \RNAf
\cite{Hofacker1994} and {\sc Mfold}~\cite{Zuker2003}. \And\ \etal\ improved
estimated values in the Turner~1999 parameter set by applying sophisticated
machine learning techniques to training 363~free parameter values
\cite{Andronescu2007} . These parameters were adjusted using a training set of
3439~reference structures and 946~thermodynamic measurements by optical
melting resulting in the \And~2007 parameter set. They observed an average
performance increase of 7\% on a test set of 1660~sequences containing several
biological classes, including tRNA, RNase~P, rRNA and SRP~RNA.

The Zuker-Stiegler algorithm traditionally considers only the change in free
energy for a given RNA~\ss\ conformation in thermodynamic equilibrium, but
does not consider the process of RNA~structure formation, \ie\ how the
RNA~sequence arrives at the MFE~structure. Rather, the Zuker-Stiegler
algorithm implicitly assumes that the input RNA~sequence (1) is already fully
synthesised, (2) in thermodynamic equilibrium and (3) will always be able to
reach the RNA~structure that minimises the overall free energy of the
molecule. We know from a range of experiments, however, that RNA~molecules
start to fold while they emerge during transcription, that they are not
necessarily in thermodynamic equilibrium during structure formation \iviv\ and
that they may get trapped on their kinetic folding pathway. That RNA~molecules
overall proceed towards the MFE~structure over time is only an approximation
of the complex reality \iviv. As the molecule emerges from the polymerase,
however, local structures immediately begin to form. Formation of long-range
base pairs may require disruption of these local structures, and their folding
rate may be prohibitively slow due to high energy barriers.  That is, the
molecule may never reach the minimum free energy structure due to kinetic
considerations.  The structure formation \iviv\ may get further complicated
due \emph{trans} interactions between the RNA~sequence and other molecules in
the living cell which we ignore for now.

We propose a new method for RNA~\ss\ prediction, \cof, that takes into account
some effects of co-transcriptional folding. The key effect that we aim to
model is that during co-transcriptional folding \iviv, it does matter to a
given sequence position whether a potential pairing partner is available for
base-pairing or not. To capture this, we model the distance along the sequence
between base pairing sequence positions. \cof\ is a modification to the
Zuker-Stiegler algorithm \cite{Zuker1981}, and it was implemented using the
\RNAf\ source code from the {\sc ViennaRNA} package \cite{Hofacker1994, Lorenz2011}.

\cof\ calculates energies in the same fashion as in \RNAf, but all energy
contributions associated with a base pair are modified by a scaling function
according to the number of nucleotides between the pair (\ie\ the distance
$d$).  This scaling function $\gamma(d)$ models the exponential decay in
reachability as function of the nucleotide distance $d$ between the two
potential pairing partners and depends on two parameters $\alpha$ and $\tau$
(\Myref{eq}{scalingfunctioneq}, \Myref{fig}{scalingfunction}). Both parameters
have a straightforward interpretation. The value of $\alpha$ specifies the
range of the scaling function (\eg\ when $\alpha$ is $0.2$, the affected free
energies will range from $80$\% to $100$\% of their original values). The
value of $\tau$ determines the rate of the exponential decay, where low values
of $\tau$ result in a steep decay function.

\begin{equation}
    \mylabel{eq}{scalingfunctioneq}
    \gamma (d) := \alpha \cdot (e^{-\frac{d}{\tau}} - 1) + 1
\end{equation}

The scaling function $\gamma(d)$ is only used in conjunction with energy
values in the $C_{i,j}$ calculation because these correspond to predicted base
pairs. The function is not applied to the energy of subsequences in order to
avoid multiple applications to the same value. The function is applied both to
elements with positive energy, such as loops and bulges, as well as to those
with negative energy, such as stacking interactions.  This is necessary to
preserve the relative magnitude of the contributions from structural
components, see \Myref{eq}{cofoldcijeq} and \Myref{fig}{cofoldijdiagrams} for
modified $C_{i,j}^{\prime}$. The $FML_{i,j}$ calculation remains the same as
in \RNAf.

\begin{equation}
  \mylabel{eq}{cofoldcijeq}
        C_{i,j}^{\prime} = min \begin{dcases*} 
                \gamma(d_{i,j}) \cdot hairpin_{i,j} & open a helix with hairpin loop \\
                min_{i<p<q<j} \{C_{p,q} + \gamma(d_{i,j}) \cdot Stack_{(i,j),(p,q)}\} & stack, bulge or internal loop \\
                min_{k,l \in {1,2}} \{ FML_{i+k, j-l} + \gamma(d_{i,j}) \cdot dangle\} & open a helix with nested substructure \\
            \end{dcases*}
\end{equation}

The output of \cof\ is an RNA \ss\ which promotes base pairs according to the
above scaling function.  The predicted RNA~\ss\ therefore captures both
thermodynamic contributions as well as effects due to co-transcriptional
structure formation. Like \RNAf, \cof\ allows the user to select a
thermodynamic parameter set and the running time of \cof\ also scales with
$O(L^{3})$, where $L$ denotes the length of the input sequence. For
performance evaluation, we use both the Turner~1999 (\cof) and the \And~2007
(\cofA) parameter sets introduced above.

\paragraph{(4) Parameter training}

\cof\ has two free parameters: $\alpha$ and $\tau$. Due to the small number of
parameters, they were trained using a simple brute force scheme. \cof\ was run
on all sequences of the combined data set and performance metrics were
calculated for each $(\alpha, \tau)$ combination in set $P$ (defined in
\Myref{eq}{parametercombinations}). The Turner~1999 thermodynamic parameter
set \cite{Mathews99} was used for $(\alpha, \tau)$ parameter training.

\begin{equation}
    \mylabel{eq}{parametercombinations}
    P := \{0.05, 0.10, \ldots, 0.90, 0.95\} \times \{40, 80, \ldots, 1160, 1200\}
\end{equation}

\begin{align}
    \mylabel{eq}{performancemetric}
    \overline{MCC}_{\alpha, \tau}^{S} :=  & \frac{\sum_{s \in S}MCC_{\alpha, \tau}^{s}}{|S|} \\
    & \text{where} \: (\alpha, \tau) \in P \: \text{and} \: S \: \text{is the sequence set} \notag
\end{align}

\begin{equation}
    \mylabel{eq}{deltaperformancemetric}
    \overline{\Delta MCC}_{\alpha, \tau}^{S} := \overline{MCC}_{\alpha, \tau}^{S} -
    \overline{MCC}_{\text{RNAfold}} 
\end{equation}

Performance metrics were found to be highly correlated in $\alpha$ and $\tau$
(\Myref{fig}{linearfit}~(right), \Myref{fig}{colourcoding}).  To demonstrate
this, linear regression was performed on the $\overline{\Delta MCC}$ matrix
(\Myref{fig}{linearfit}~(left)). We first compiled a set of triples $Q =
\{(\alpha, \tau, \overline{\Delta MCC}_{\alpha, \tau})\}$, for which
$\overline{\Delta MCC}_{\alpha, \tau}$ is in the 97$^{th}$ quantile of the
performance matrix.  Weighted linear regression was performed with $\alpha$
and $\tau$ as dimensions, and $\overline{\Delta MCC}$ as the weight. The
regression line fits the data with an $R^{2}$ value of $98.4$\%, indicating
that variability in $\tau$ highly accounts for the variability in
$\alpha$. Regression line (solid) and its $95$\% confidence region (dotted)
are plotted in \Myref{fig}{linearfit} (left).

Twenty trials of five-fold cross validation were performed to determine
robustness of parameter training.  In each trial, the combined data set $D$ is
randomly divided into five partitions $D_i$. For each partition, the optimal
parameter combination is determined for the remaining sequences
(\Myref{eq}{crossvalidationeq}). The cross validation results are plotted in
\Myref{fig}{linearfit}~(right), where the integer in each cell indicates the
number of trials where that parameter combination was optimal. The optimal
parameter values highly cluster around the linear regression line show in
\Myref{fig}{linearfit}~(left).

\begin{align}
    \mylabel{eq}{crossvalidationeq}
    (\alpha_{opt}, \tau_{opt}) := & (\alpha, \tau) \; s.t. \; \overline{\Delta MCC}_{\alpha, \tau}^{T} = \text{max}(\overline{\Delta
    MCC}^{T}) \\
    & \text{where} \; T := D \setminus D_i  \notag
\end{align}

The default parameter combination for \cof\ is $\alpha = 0.5, \tau=640$. This
parameter set maximises $\overline{MCC}$ for the combined dataset. The default
parameter combination is marked with an "X" in \Myref{fig}{linearfit}~(left)
which shows that it lies directly on the linear regression line.

\paragraph{(5) Calculation of free energy differences}

We define $\Delta \Delta G$ as the difference between the free energy ($\Delta
G$) of a given prediction and the corresponding \RNAf\ prediction. We
calculate these values for \RNAfA, \cof, and \cofA. Because the \And~2007
parameters use modified the free energy values, we use {\sc RNAeval} from the
{\sc ViennaRNA} package~\cite{Hofacker1994, Lorenz2011} to calculate the free energy
of each predicted structure on an equal footing. Unlike \RNAf\ which predicts
a minimum free energy structure from a sequence, {\sc RNAeval} calculated the
free energy for an input RNA~structure according to the provided thermodynamic
parameters. For consistency, we use the parameters Turner~1999 thermodynamic
model~\cite{Mathews99} for all $\Delta \Delta G$ calculation. For a prediction
program $X$ which corresponds to \RNAfA\, \cof\ or \cofA, we define $\Delta
\Delta G_X$ and $\% \Delta \Delta G_X$ as follows.

\begin{equation}
    \Delta \Delta G_X = \Delta G_X - \Delta G_{\text{RNAfold}}
    \mylabel{eq}{deltadeltag}
\end{equation}

\begin{equation}
    \% \Delta \Delta G_X = 100 \cdot \frac{\left( \Delta G_X -
    \Delta G_{\text{RNAfold}} \right)}{|\Delta G_{\text{RNAfold}}|}
    \mylabel{eq}{pdeltadeltag}
\end{equation}

\vfill

\pagebreak

\section*{Tables}

\begin{table}[h!]
\centering
\begin{tabular}{l|r|r|r}
            & \multicolumn{1}{c}{long data set} & \multicolumn{2}{|c}{combined data set} \\ \hline
clade       & $>$ 1000~nt              & all   & $\le$ 1000~nt      \\ \hline
Bacteria    & 15                       & 69    &   (54)             \\
Eukaryotes  & 15                       & 112   &   (97)             \\
Virus       & 0                        & 20    &   (20)             \\
Archea      & 17                       & 33    &   (16)             \\
Chloroplast & 14                       & 14    &    (0)             \\ \hline
sum         & 61                       & 248   &   (187)            \\[1em]
sequence length (nt) &                      &       &                    \\ \hline
average & 2397             & 776   &   (247)            \\
minimum & 1245             & 110   &   (110)            \\
maximum & 3578             & 3578  &   (628)            \\ 
\end{tabular}
\caption{{\bf Evolutionary composition and length statistics for the long and the combined data set.} Numbers in brackets specify the respective numbers for the short sequences 
in the combined data set.}
\mylabel{tab}{Dataset}
\end{table}

\begin{table}[H]
\centering
\begin{tabular}{l|rrrrrrl}
biological class                               & A.len &   clade &   N.seq &   N.ext &   max    &  source	& ID              \\
                                               &       &         &         &         &   ppid     &             &                 \\ 
                                               & (nt)  &         &         &         &   (\%)     &             &                 \\ \hline
16S rRNA (archea)                              & 1545  &   A     &   40    &   8     &   85     &  \CRW\	& 16S archaea     \\ 
23S rRNA (archea)                              & 3153  &   A     &   40    &   9     &   85     &  \CRW\	& 23S archaea     \\ 
16S rRNA (bacteria)                            & 1661  &   B     &   144   &   7     &   70      &  \CRW\	& 16S bacteria    \\ 
23S rRNA (bacteria)                            & 3046  &   B     &   40    &   8     &   85     &  \CRW\	& 23S bacteria    \\ 
16S rRNA (chloroplast)                         & 1558  &   C     &   40    &   5     &   85     &  \CRW\	& 16S chloroplast \\ 
23S rRNA (chloroplast)                         & 3722  &   C     &   40    &   9     &   80      &  \CRW\	& 23S chloroplast \\ 
16S rRNA (eukaryote)                           & 1867  &   E     &   40    &   7     &   85     &  \CRW\	& 16S eukaryote   \\ 
23S rRNA (eukaryote)                           & 4105  &   E     &   40    &   8     &   85     &  \CRW\	& 23S eukaryote   \\ \hline
snRNA                                          &   184 &   E     &   87    &   14    &   80      &  \RFAM\	& RF00003         \\ 
U2 spliceosomal RNA                            &   270 &   E     &   181   &   10    &   50      &  \RFAM\	& RF00004         \\ 
Nuclear RNase P                                &   622 &   E     &   77    &   11    &   45     &  \RFAM\	& RF00009         \\ 
snoRNA                                         &   236 &   E     &   14    &   9     &   85     &  \RFAM\	& RF01256         \\ 
snoRNA                                         &   394 &   E     &   4     &   1     &   85     &  \RFAM\	& RF01267         \\ 
snoRNA                                         &   373 &   E     &   18    &   9     &   85     &  \RFAM\	& RF01296         \\ 
U4 spliceosomal RNA                            &   273 &   E     &   160   &   11    &   50      &  \RFAM\	& RF00015         \\ 
U5 spliceosomal RNA                            &   178 &   E     &   153   &   9     &   45     &  \RFAM\	& RF00020         \\ 
ciliate telomerase RNA comp.\                  &   270 &   E     &   19    &   11    &   80      &  \RFAM\	& RF00025         \\ 
RNase MRP                                      &   903 &   E     &   40    &   12    &   50      &  \RFAM\	& RF00030         \\ 
RNase P                                        &   511 &   B     &   88    &   8     &   60      &  \RFAM\	& RF00011         \\ 
CsrB RNA                                       &   391 &   B     &   11    &   7     &   85     &  \RFAM\	& RF00018         \\ 
lysine riboswitch                              &   232 &   B     &   37    &   14    &   65     &  \RFAM\	& RF00168         \\ 
Mg riboswitch - Ykok leader                    &   216 &   B     &   85    &   14    &   65     &  \RFAM\	& RF00380         \\ 
Ornate extremophilic RNA                       &   676 &   B     &   7     &   6     &   85     &  \RFAM\	& RF01071         \\ 
Pestivirus IRES                                &   286 &   V     &   23    &   5     &   85     &  \RFAM\	& RF00209         \\ 
Tombusvirus 5' UTR                             &   180 &   V     &   7     &   5     &   85     &  \RFAM\	& RF00171         \\ 
Aphthovirus IRES                               &   471 &   V     &   87    &   4     &   85     &  \RFAM\	& RF00210         \\ 
Cripavirus IRES                                &   208 &   V     &   6     &   6     &   80      &  \RFAM\	& RF00458         \\ 
tRNA-like structures                           &   137 &   V     &   5     &   5     &   80      &  \RFAM\	& RF01084         \\ 
Archaeal RNase P                               &   533 &   A     &   25    &   16    &   80      &  \RFAM\	& RF00373         \\ 
\end{tabular}
\caption{{\bf RNA~families of the long and the combined data set.} All sequences of the long data set derive from alignments of the \CRW\ data base (top), whereas
  the short sequences from the combined data set all derive from alignments of the \RFAM\ data base (bottom). For each original alignment from either data base, 
  \ie\ each row in this table, we specify the alignment length in nucleotides (A.len), 
  the evolutionary origin of its sequences (clade, A - Archea, B - Bacteria, C - Chloroplast, V - Virus, E - Eukaryotes), the number of sequences (N.seq), data base (source) and 
  identifier in that data base (ID). We also specify, for each original alignment, how many sequences we extracted (N.ext) and what their maximum pairwise percent identify is in terms
  of primary sequence conservation (max.~ppid). IRES stands for internal ribosomal entry site.}
\mylabel{tab}{Dataset2}
\end{table}

\begin{table}[H]
\centering
\begin{tabular}{l|rrrrlrrrrr}
alignment (ID)  & A.len & av.\ seq.\ & av.\  & gaps    & n                 & bpairs & canonical & covar.\ \\
                & (nt)  & length    & ppid  & (\%)    & (\%)              &        & bpairs    &        \\
                &       & (nt)      & (\%)  &         &                   &        & (\%)      &        \\ \hline
16S archaea     & 1545  & 1477      & 81.8 &   4.4    & $5 \cdot 10^{-7}$  &  458   & 95.2      &   0.343 \\     
23S archaea     & 3153  & 2945      & 74.9 &   6.6    & $6 \cdot 10^{-7}$  &  852   & 95.0      &   0.408 \\     
16S bacteria    & 1661  & 1520      & 76.7 &   8.5    & $2 \cdot 10^{-2}$  &  453   & 93.4      &   0.352 \\     
23S bacteria    & 3046  & 2904      & 79.2 &   4.6    & $6 \cdot 10^{-7}$  &  868   & 94.3      &   0.358 \\     
16S chloroplast & 1558  & 1490      & 90.2 &   4.4    & $5 \cdot 10^{-7}$  &  440   & 93.9      &   0.113 \\     
23S chloroplast & 3722  & 2941      & 74.8 &  21.0    & $0              $ &  869   & 90.1      &   0.253\\     
16S eukaryote   & 1867  & 1708      & 73.3 &   8.5    & $1 \cdot 10^{-7}$  &  370   & 84.3      &   0.162 \\     
23S eukaryote   & 4105  & 3476      & 78.7 &  15.3    & $1 \cdot 10^{-7}$  &  998   & 88.1      &   0.084 \\ \hline
RF00003	        &   184 &  162      & 63.1 &  11.8    & $0	     $    &   40   & 93.2      &   0.493      \\	 
RF00004	        &   270 &  193      & 58.4 &  28.4    & $1 \cdot 10^{-2}$  &   45   & 92.8      &   0.496 \\	 
RF00009	        &   622 &  315      & 40.7 &  49.3    & $8 \cdot 10^{-5}$  &   62   & 89.3      &   0.397 \\	 
RF01256	        &   236 &  208      & 60.6 &  11.9    & $0	      $   &   54   & 92.7      &   0.457      \\	 
RF01267	        &   394 &  384      & 72.5 &   2.6    & $0	      $   &  128   & 94.3      &   0.295      \\	 
RF01296	        &   373 &  325      & 63.7 &  13.0    & $0	      $   &   57   & 91.4      &   0.339      \\	 
RF00015	        &   273 &  147      & 52.2 &  46.2    & $5 \cdot 10^{-5}$  &   31   & 91.5      &   0.604 \\	 
RF00020	        &   178 &  117      & 51.7 &  34.1    & $4 \cdot 10^{-5}$  &   30   & 94.0      &   0.694 \\	 
RF00025	        &   270 &  186      & 42.5 &  31.1    & $0	      $   &   39   & 86.4      &   0.395      \\	 
RF00030	        &   903 &  303      & 34.7 &  66.5    & $0	      $   &   74   & 88.3      &   0.470       \\	 
RF00011	        &   511 &  373      & 63.0 &  27.1    & $0	      $   &  105   & 95.1      &   0.500        \\	 
RF00018	        &   391 &  350      & 62.4 &  10.5    & $0	      $   &   49   & 96.8      &   0.368      \\	 
RF00168	        &   232 &  183      & 46.1 &  21.2    & $0	      $   &   53   & 90.3      &   0.580       \\	 
RF00380	        &   216 &  170      & 59.6 &  21.4    & $0	      $   &   47   & 94.5      &   0.471      \\	 
RF01071	        &   676 &  609      & 59.9 &   9.9    & $0	      $   &  159   & 90.2      &   0.378      \\	 
RF00209	        &   286 &  275      & 89.2 &   3.9    & $0	      $   &   75   & 98.8      &   0.191      \\	 
RF00171	        &   180 &  159      & 67.3 &  11.4    & $0	      $   &   34   & 97.9      &   0.403      \\	 
RF00210	        &   471 &  461      & 85.4 &   2.1    & $0	      $   &  122   & 98.3      &   0.181      \\	 
RF00458	        &   208 &  201      & 55.4 &   3.5    & $0	      $   &   60   & 95.0      &   0.757      \\	 
RF01084	        &   137 &  128      & 51.0 &   6.9    & $0	      $   &   43   & 97.2      &   0.795      \\	 
RF00373	        &   533 &  311      & 49.4 &  41.7    & $0	      $   &   87   & 90.0      &   0.537      \\	 
\end{tabular}
\caption{{\bf Alignment quality and phylogenetic support for the reference RNA \ss s.} For each original alignment, \ie\ each row in this table, we
  specify the alignment length in nucleotides (A.len), the average length of each non-gapped sequence in that alignment (av.\ seq.\ length), 
  the average pairwise percent identity between pairs of sequences in the alignment in terms of primary sequence conservation (av.\ ppid), 
  the average fraction of gaps per sequence in the alignment (gaps), the average fraction of ambiguous (not A,C,G,T,U,-) nucleotide symbols per sequence 
  in the alignment (n), the number of base pairs in the reference RNA \ss\ for that alignment (bpairs), the average fraction of sequences in the alignment 
  that have a consensus base-pair per conserved base-pair of the reference \ss\ (canonical bpairs) and the covariation (covar.) as defined in \RFAM\ \cite{Griffiths-Jones2005} which 
  measures how well the base pairs of the reference RNA \ss\ are supported by co-variation (high means good).
}
\mylabel{tab}{Dataset3}
\end{table}

\begin{table}[H]
\centering
\begin{tabular}{l|rrrr}\hline
 \multicolumn{5}{c}{long data set}                     \\ \hline 
          & TPR (\%) & FPR (\%) &  PPV (\%) & MCC (\%) \\ \hline
\RNAf\    & 46.30    & 0.0176   &  39.74    & 42.81    \\  
\RNAfA\   & 52.02    & 0.0160   &  44.76    & 48.17    \\
\cof\     & 52.83    & 0.0159   &  45.79    & 49.10    \\
\cofA\    & 57.80    & 0.0145   &  50.06    & 53.70    \\ \hline
 \multicolumn{5}{c}{combined data set}                 \\ \hline 
          & TPR (\%) & FPR (\%) &  PPV (\%) & MCC (\%) \\ \hline
\RNAf\    & 57.87    & 0.1132   &  45.27    & 50.86    \\
\RNAfA\   & 58.98    & 0.1152   &  46.16    & 51.83    \\
\cof\     & 60.38    & 0.1097   &  47.56    & 53.26    \\
\cofA\    & 61.51    & 0.1112   &  48.42    & 54.22    \\ \hline
 \multicolumn{5}{c}{short sequences only}              \\ \hline 
          & TPR (\%) & FPR (\%) &  PPV (\%) & MCC (\%) \\ \hline
\RNAf\    & 61.64    & 0.1444   &  47.08    & 53.48    \\
\RNAfA\   & 61.25    & 0.1475   &  46.61    & 53.02    \\
\cof\     & 62.84    & 0.1403   &  48.14    & 54.62    \\
\cofA\    & 62.72    & 0.1428   &  47.88    & 54.39    \\    
\end{tabular}
\caption{{\bf \cof\ predictive power for base pairs for all data sets.} 
  The performance accuracy of \cof\ and \cofA\ compared to \RNAf\ and \RNAfA\ for the test
  set as measured in terms of true positive rate ($TPR = 100 \cdot TP/(TP + FN)$), 
  false positive rate ($ FPR = 100 \cdot FP/(FP + TN) $),
  positive predictive value ($ PPV = 100 \cdot TP/(TP + FP)$) and Matthew's correlation coefficient 
  ($ MCC = 100 \cdot (TP \cdot TN - FP \cdot FN)/\sqrt{(TP + FP) \cdot (TP + FN) \cdot (TN + FP) \cdot (TN + FN)}$),
  where TP denotes the numbers of true positives, TN the true negatives, 
  FP the false positives and FN the false negatives.}
\mylabel{tab}{Performanceall}
\end{table}

%
%
%
%
\begin{table}[H]
\centering
\begin{tabular}{l|rrr|rrr|rrr}
               \multicolumn{10}{c}{Summary of \% $\Delta\Delta$G distributions} \\
            & \multicolumn{3}{c}{long data set} & \multicolumn{6}{|c}{combined data set} \\ \hline
            & \multicolumn{3}{c}{$>$ 1000~nt}   & \multicolumn{3}{|c}{all}     & \multicolumn{3}{|c}{$\le$ 1000~nt} \\
            & av.\ $\pm$ stdev & min & max      & av.\ $\pm$ stdev & min & max & av.\ $\pm$ stdev & min & max   \\ \hline
\RNAfA\     & $4.7 \pm  1.9$ &  1.4  &   11.1   & $5.0 \pm  3.5$ &  -2.3 &15.4 & $5.1 \pm  3.9$ &  -2.3 &   15.4 \\
\cof\       & $1.8 \pm  1.0$ &  0.2  &    4.4   & $0.5 \pm  1.2$ &  -5.0 & 4.4 & $0.1 \pm  0.9$ &  -5.0 &    3.8 \\
\cofA\      & $6.8 \pm  2.4$ &  1.7  &   13.1   & $5.9 \pm  3.8$ &  -2.3 &18.2 & $5.6 \pm  4.1$ &  -2.3 &   18.2 \\
\end{tabular}
\caption{{\bf Summary of relative free energy difference distributions of predicted structures w.r.t.\ the
  MFE~structured predicted by \RNAf\ for the same input sequences, for all data sets.}
}
\mylabel{tab}{deltaGdistributions}
\end{table}

\begin{table}[H]
\centering
\begin{tabular}{l|rrr}
   \multicolumn{4}{c}{Linear fit to $\Delta$ MCC versus \% $\Delta\Delta$G distributions} \\
            & \multicolumn{3}{|c}{long data set} \\ \hline
            & \multicolumn{3}{|c}{$>$ 1000~nt} \\ 
            & intercept $\pm$ stdev  & slope $\pm$ stdev & R$^2$ (\%)  \\ 
\RNAfA\     & $7.0 \pm 2.4$ & $ -0.34 \pm 0.48  $ & $0.85$ \\
\cof\       & $3.5 \pm 1.6$ & $  1.52 \pm 0.78  $ & $6.06$ \\
\cofA\      & $9.2 \pm 3.1$ & $  0.25 \pm 0.43  $ & $0.56$ \\ \hline
            & \multicolumn{3}{|c}{combined data set} \\ 
            & intercept $\pm$ stdev  & slope $\pm$ stdev & R$^2$ (\%) \\ \hline
\RNAfA\     & $1.0 \pm 1.4$ & $  0.0008 \pm 0.23$ & $5.6 \cdot 10^{-06}$ \\
\cof\       & $2.1 \pm 0.6$ & $  0.59   \pm 0.47$ & $0.64$ \\
\cofA\      & $2.1 \pm 1.6$ & $  0.21   \pm 0.23$ & $0.34$ \\ \hline
            & \multicolumn{3}{|c}{short sequences only} \\ 
            & \multicolumn{3}{|c}{$\le$ 1000~nt} \\ 
            & intercept $\pm$ stdev  & slope $\pm$ stdev & R$^2$ (\%) \\ \hline
\RNAfA\     & $-0.8 \pm 1.6$ & $  0.06  \pm 0.25$ & $0.03$ \\
\cof\       & $ 1.3 \pm 0.7$ & $ -2.21  \pm 0.75$ & $4.44$ \\
\cofA\      & $ 0.7 \pm 1.7$ & $  0.03  \pm 0.25$ & $0.01$ \\
\end{tabular}
\caption{{\bf Details of the linear fits to the $\Delta$ MCC versus \% $\Delta\Delta$G distributions.} }
\mylabel{tab}{deltadeltaGversusdeltaMCCfits}
\end{table}

\pagebreak

\section*{Figures}

\begin{figure}[H]
  \begin{center}
    \includegraphics[width=8cm]{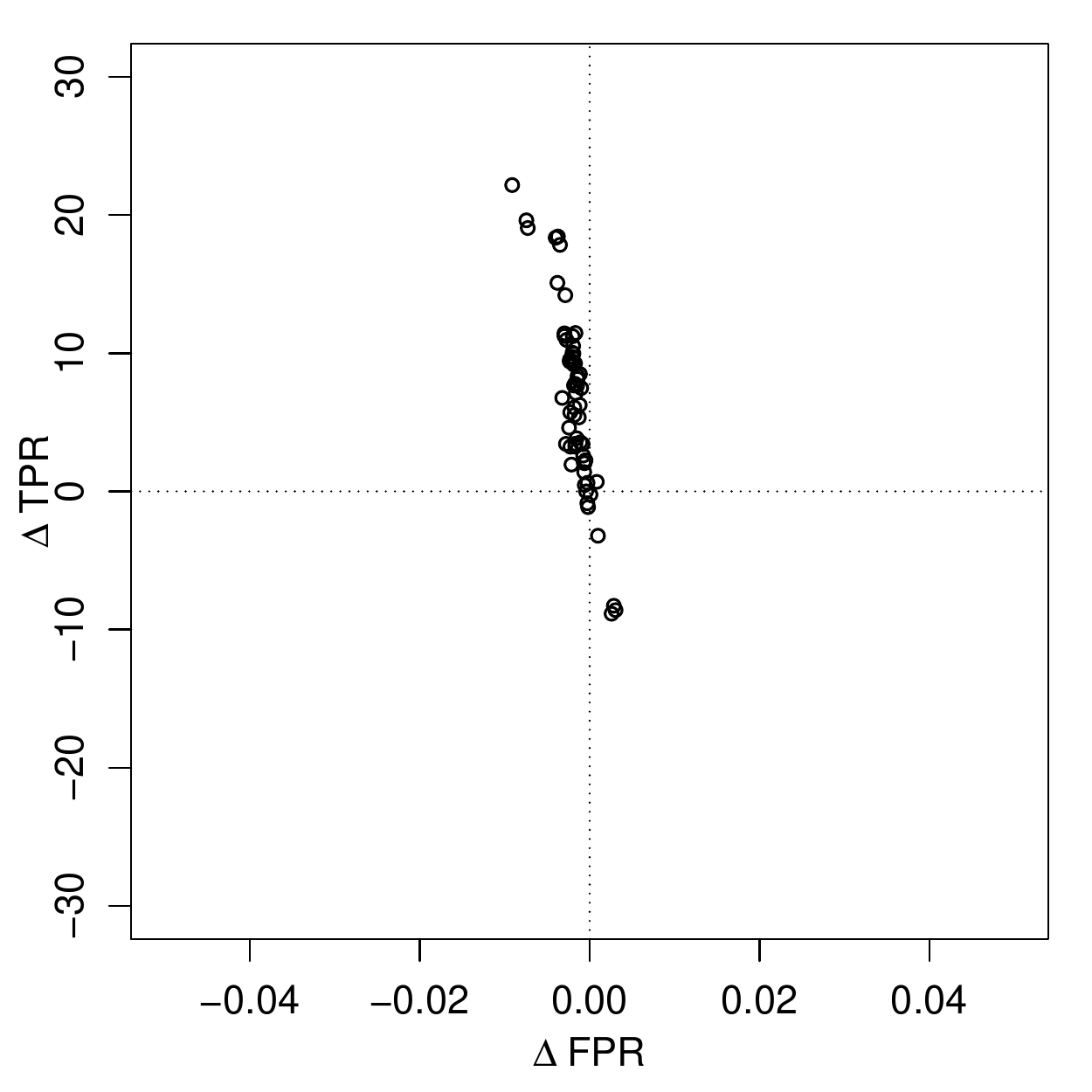}
  \end{center}
  \caption{ {\bf Changes in prediction accuracy for the structures predicted
    by \cof\ for all sequences of the long data set.}  We report the
    prediction accuracy for base pairs in terms of relative changes in
    true positive rate ($TPR = 100 \cdot TP/(TP + FN)$) and false positive rate ($ FPR = 100 \cdot
    FP/(FP + TN) $) by comparing the prediction accuracy of the structures
    predicted by \cof\ to those predicted by \RNAf. TP denotes the numbers of
    true positives, TN the true negatives, FP the false positives and FN the
    false negatives.}
  \mylabel{fig}{performanceTPRversusFPR}
\end{figure}

\begin{figure}[H]
  \begin{center}
    \includegraphics[width=7cm]{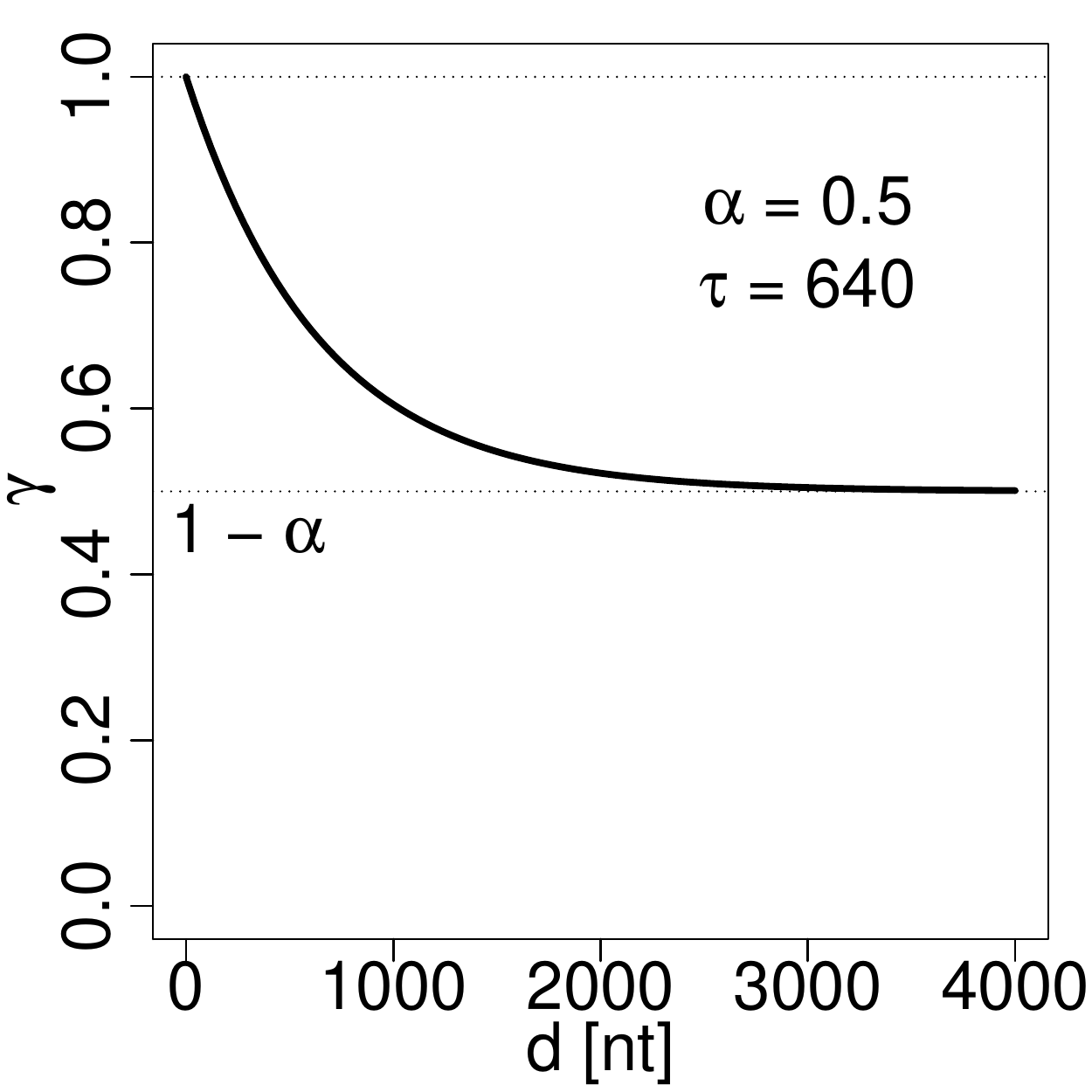}
  \end{center}
  \caption{{\bf Scaling function $\gamma$ of \cof.}}
  \mylabel{fig}{scalingfunction}
\end{figure}

\begin{figure}[H]
  \begin{center}
    \includegraphics[width=7cm]{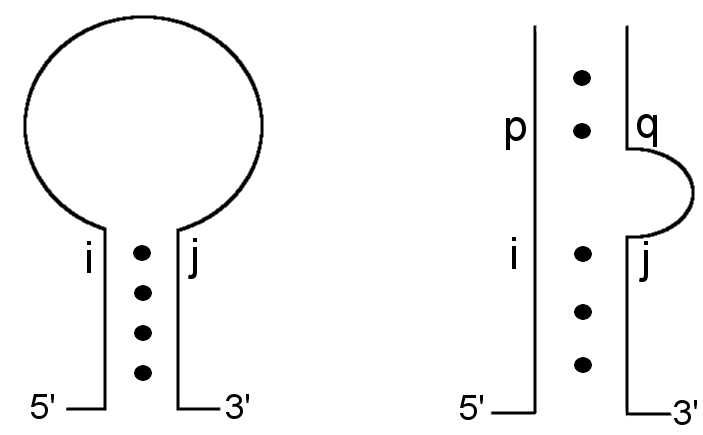}
  \end{center}
  \caption{{\bf Details of the sequence coordinates affected by the scaling
      function of \cof\ in Equation~(13).}}
  \mylabel{fig}{cofoldijdiagrams}
\end{figure}

\begin{samepage}

\begin{figure}[H]
  \begin{center}
    \includegraphics[width=16cm]{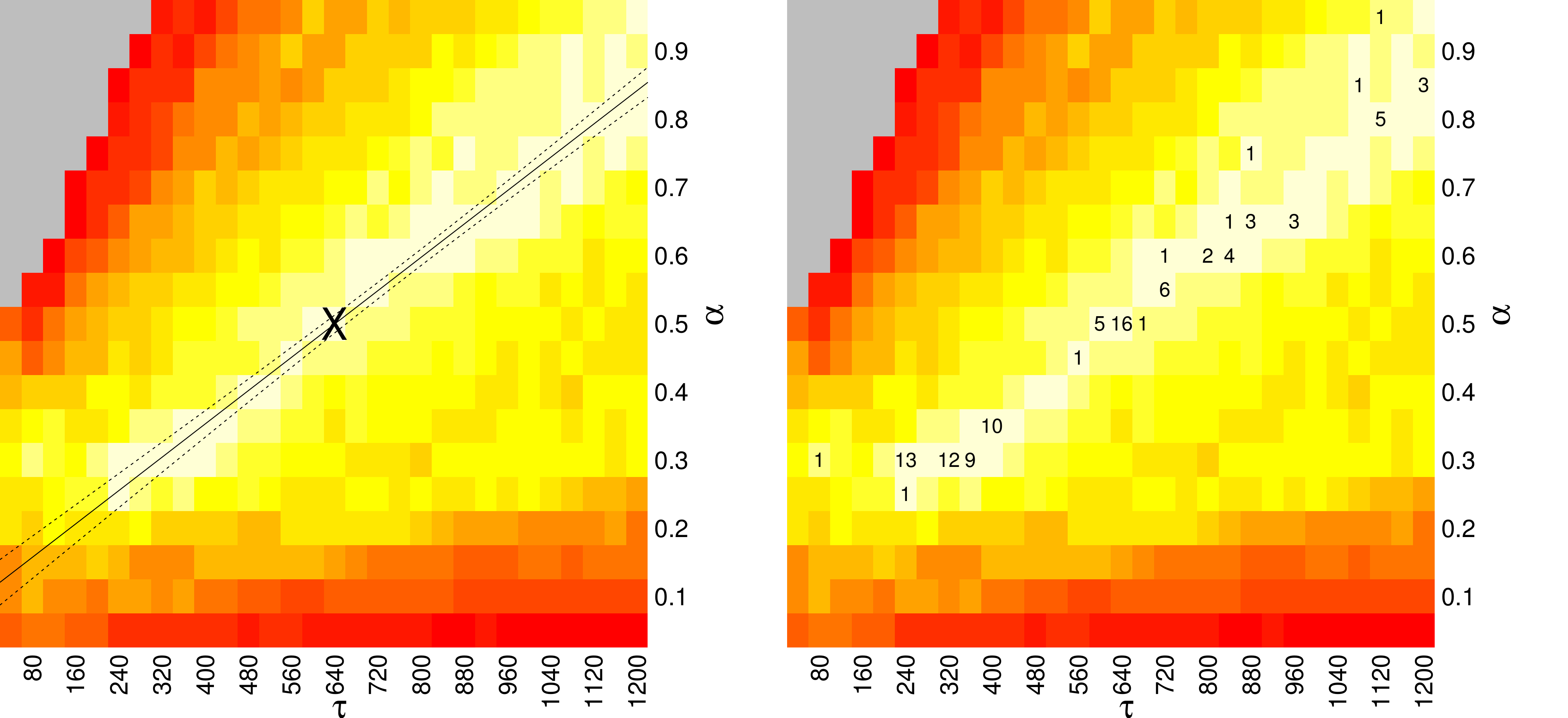}
  \end{center}
  \caption{{\bf Training of parameters in \cof: linear fit and robustness.}
    Left figure, heatmap showing the average MCC differences w.r.t.\ \RNAf\ as
    function of the $\tau$ (x-axis) and $\alpha$ (y-axis) parameters values.
    The average MCC differences are indicated via the colours from high
    (bright yellow) to low (dark red), see Figure~9 for details.  The solid
    line corresponds to the linear regression line ($\alpha = a \cdot \tau +
    b$ with a slope of $a = 6.1 \cdot 10^{-4} \pm 2 \cdot 10^{-5}$ and an
    intercept of $b = 0.105 \pm 0.016$). The two dotted lines delineate the
    95\% confidence region. The asterisk shows parameter pair with highest
    average MCC ($\alpha = 0.50$ and $\tau = 640$) which is the parameter
    combination used in \cof\ and \cofA.  Right figure, same heatmap as in
    left figure, but this time showing the count of trials in 20~trials of
    five-fold cross-validation where that the corresponding pair of parameter
    values has the highest average MCC for the set of training sequences.}
\mylabel{fig}{linearfit}
\end{figure}

\begin{figure}[H]
  \begin{center}
    \includegraphics[width=6cm]{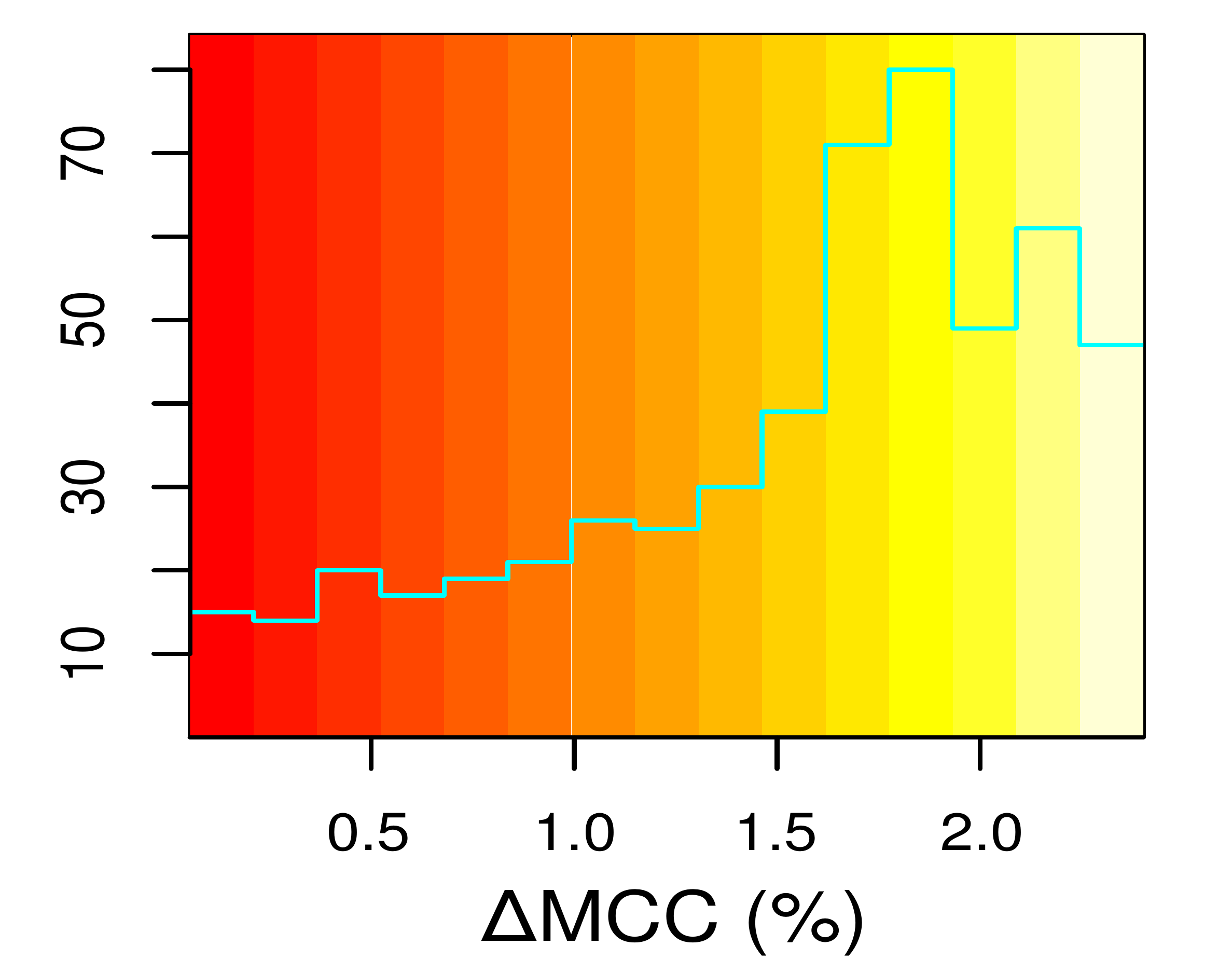}
  \end{center}
  \caption{{\bf Colour coding for the heatmaps for parameter training of
      \cof.}  Histogram of the average MCC differences of \cof\ w.r.t.\ \RNAf,
    ranging from light yellow for the largest improvement of performance
    accuracy to dark red for a small performance improvement. A grey colour in
    Figure~8 corresponds to an improvement accuracy that is smaller than the
    range covered in this histogram.}
  \mylabel{fig}{colourcoding}
\end{figure}

\end{samepage}

\begin{figure}[H]
  \begin{center}
    \includegraphics[width=16cm]{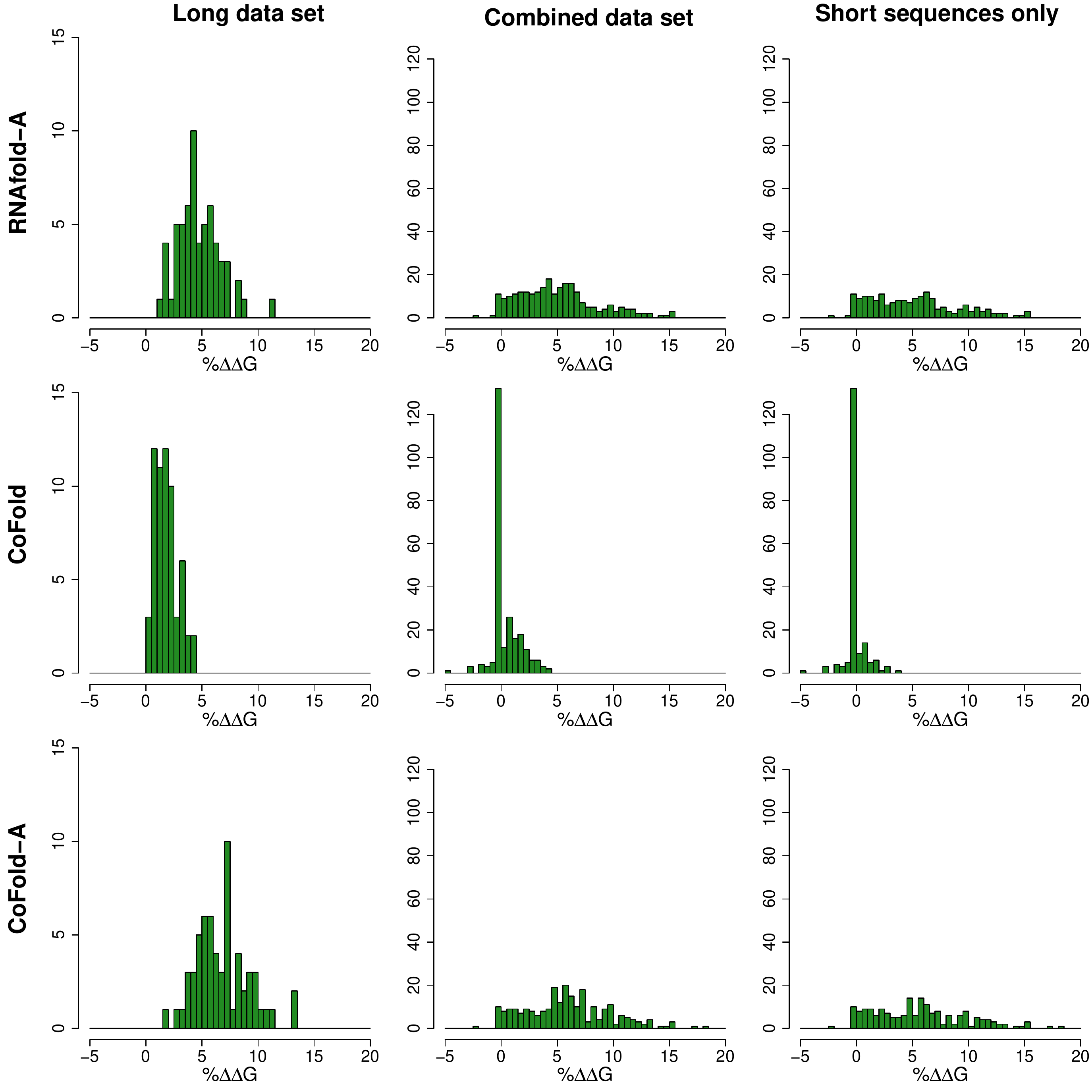}
  \end{center}
  \caption{{\bf Relative free energy difference distributions of the predicted
      structures w.r.t.\ the MFE~structures predicted by \RNAf\ for the same
      input sequences, for all data sets.}  Results for the long data set
    (left column), the combined data set (middle column) and the short
    sequences of the combined data set (right column).  For each data set,
    three histrograms show the relative free energy differences of the
    RNA~structures predicted by \RNAfA\ w.r.t. the MFE~structures predicted by
    \RNAf\ for the same sequence (top row), of the RNA~structures predicted by
    \cof\ w.r.t.\ the MFE~structures predicted by \RNAf\ (middle row) and of
    the RNA~structures predicted by \cofA\ w.r.t\ the MFE~structures predicted
    by \RNAfA\ (bottom row). The free energies of all structures are
    calculated using the Turner~1999 energy parameters.}
  \mylabel{fig}{deltaGdistributionshistogram}
\end{figure}

\begin{figure}[H]
  \begin{center}
    \includegraphics[width=16cm]{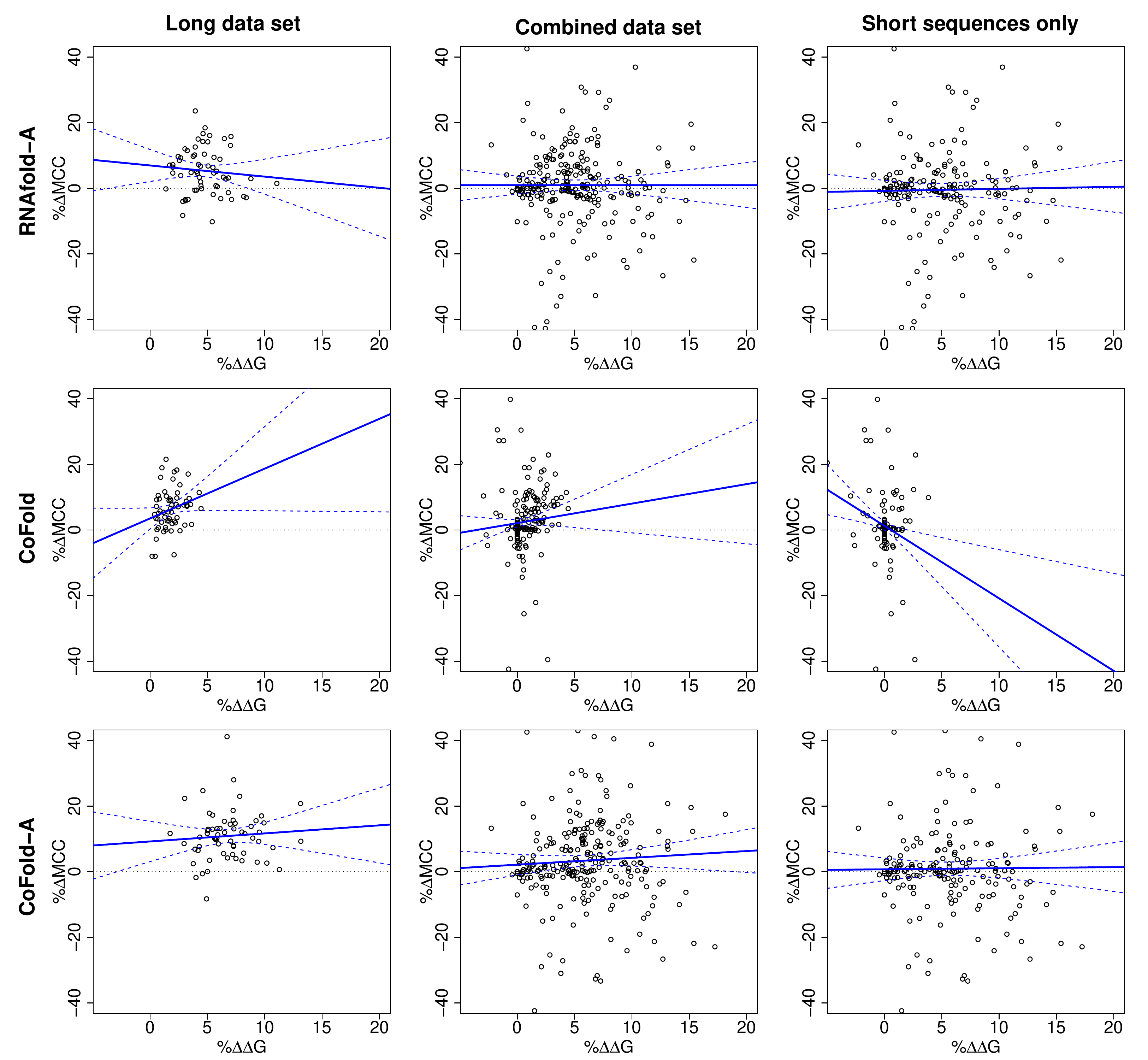}
  \end{center}
  \caption{{\bf Differences in prediction accuracy versus relative free energy
      changes of the predicted structures w.r.t.\ the MFE~structures predicted
      by \RNAf\ for the same input sequences, for all data sets.}  Results for
    the long data set (left column), the combined data set (middle column) and
    the short sequences of the combined data set (right column).  For each
    data set, three figures show the change in performance accuracy in terms
    of MCC versus the relative change of free energy for the structures
    predicted by \RNAfA\ (top row) w.r.t.\ the RNA structures predicted by
    \RNAf\ for the same sequence, for the structures predicted by \cof\
    (middle row) w.r.t.\ the RNA structures predicted by \RNAf\ for the same
    sequence and for the structures predicted by \cofA\ (bottom row) w.r.t.\
    the RNA structures predicted by \RNAf\ for the same sequence.  The free
    energies of all structures are calculated using the Turner~1999 energy
    parameters.}
  \mylabel{fig}{deltadeltaGversusdeltaMCCfits}
\end{figure}

\end{document}